\documentclass[twocolumns]{aa}

\usepackage{xcolor}
\usepackage[varg]{txfonts}
\usepackage{amsmath}
\usepackage{lscape}
\usepackage{multirow}
\usepackage{pdflscape}
\usepackage{hyperref} 
\usepackage{adjustbox}
\usepackage{siunitx}

\def\kms{\hbox{km$\;$s$^{-1}$}}
\def\cm3{\hbox{cm$^{-3}$}}
\def\deg{\hbox{$^{\circ}$}}

\graphicspath{{./}{figures/}}

\begin{document}

\title{Whistler waves generated inside magnetic dips in the young solar wind: observations of the Search-Coil Magnetometer on board Parker Solar Probe}

\titlerunning{Whistler wave parameters in the young solar wind}
\author{C. Froment\inst{1}
\and O.V. Agapitov\inst{2}
\and V. Krasnoselskikh\inst{1,2}
\and S. Karbashewski\inst{2}
\and T. Dudok de Wit\inst{1,3}
\and A. Larosa\inst{4}
\and L. Colomban\inst{1}
\and D. Malaspina\inst{5,6}
\and M. Kretzschmar\inst{1}
\and V. K. Jagarlamudi\inst{7}
\and S. D. Bale\inst{2,8}
\and J. W. Bonnell\inst{2}
\and F. S. Mozer\inst{2,8}
\and M. Pulupa\inst{2}}

\institute{LPC2E, CNRS/University of Orl\'eans/CNES, 3A avenue de la Recherche Scientifique, Orl\'eans, France\\
\email{clara.froment@cnrs-orleans.fr}
\and
Space Sciences Laboratory, University of California, Berkeley, CA 94720-7450, USA
\and
International Space Science Institute, ISSI, Bern, Switzerland
\and
Department of Physics and Astronomy, Queen Mary University of London, London, UK
\and
Laboratory for Atmospheric and Space Physics, University of Colorado, Boulder, CO 80303, USA
\and
Astrophysical and Planetary Sciences Department, University of Colorado, Boulder, CO 80303, USA
\and
Johns Hopkins University Applied Physics Laboratory, Laurel, MD 20723, USA
\and
Physics Department, University of California, Berkeley, CA,  United States}

\date{Received / Accepted}

\abstract
{Whistler waves are electromagnetic waves produced by electron-driven instabilities, that in turn can reshape the electron distributions via wave-particle interactions. In the solar wind, they are one of the main candidates for explaining the scattering of the strahl electron population into the halo at increasing radial distances from the Sun and for subsequently regulating the solar wind heat flux. However, it is unclear what type of instability dominates to drive whistlers in the solar wind.}
{Our goal is to study whistler wave parameters in the young solar wind sampled by Parker Solar Probe (PSP). The wave normal angle (WNA) in particular is a key parameter to discriminate between the generation mechanisms of these waves.}
{We analyze the cross-spectral matrices of magnetic field fluctuations measured by the Search-Coil Magnetometer (SCM) and processed by the Digital Fields Board (DFB) from the FIELDS suite during PSP's first perihelion.}
{Among the 2701 wave packets detected in the cross spectra, namely individual bins in time and frequency, most were quasi-parallel to the background magnetic field but a significant part (3\%) of observed waves had oblique (> 45°) WNA. The validation analysis conducted with the time-series waveforms reveal that this percentage is a lower limit. Moreover, we find that about 64\% of the whistler waves detected in the spectra are associated with at least one magnetic dip.} 
{We conclude that magnetic dips provides favorable conditions for the generation of whistler waves. We hypothesize that the whistlers detected in magnetic dips are locally generated by the thermal anisotropy as quasi-parallel and can gain obliqueness during their propagation. We finally discuss the implication of our results for the scattering of the strahl in the solar wind.}

\keywords{Sun: heliosphere, Sun: solar wind, Waves, Plasmas}

\authorrunning{Froment et al.}
\maketitle
\section{Motivations}\label{sec:intro}

Whistler waves are circularly polarized electromagnetic waves at kinetic scales that dominate the frequency band bounded by the lower hybrid frequency $f_{LH}$ and the electron cyclotron frequency $f_{ce}$. Whistlers have extensively been studied in the near-Earth and planetary environments: Earth's ionosphere \citep{helliwell_whistlers_1965}, planetary magnetospheres \citep[e.g.][]{gurnett_whistlers_1990, horne_wave_2005, millan_review_2007, thorne_radiation_2010, 2016SSRv..200..261A, li_global_2020} and solar wind at 1~AU \citep[e.g.][]{zhang_bursts_1998,lacombe_whistler_2014, kajdic_suprathermal_2016, stansby_experimental_2016, 2019ApJ...878...41T} and are of particular interest in the context of wave-particle interactions. They can be created by different types of electromagnetic instabilities that are driven by the electron distributions \citep[see e.g.][for a review on electron-driven instabilities in the solar wind]{verscharen_electron-driven_2022}. Via wave-particle interactions they in turn can shape the electron distributions. 

In the solar wind context, whistlers are the prime candidate for explaining the modification of the electron velocity distribution function (eVDF) through the heliosphere. Recent large statistics on whistler waves were conducted at 1~AU using mainly electric field waveforms from STEREO \citep{cattell_narrowband_2020-1} of high amplitude nearly-electrostatic whistlers \citep{breneman_observations_2010}, and outside of the near-Earth environment using magnetic field measurements from HELIOS down to 0.3 AU \citep{jagarlamudi_whistler_2020}. These studies are reaffirming the interest for a global understanding of the role of whistler waves in shaping the electron distribution in the heliosphere.
The era of Parker Solar Probe \citep[PSP;][]{fox15} and Solar Orbiter \citep{muller_solar_2020} observations, now opens the possibility of extensive statistical studies of kinetic properties of the young solar wind, and more specifically to study whistler waves and the concurrent modifications in the eVDF in the young solar wind.

The solar wind eVDF is composed of three main different parts: a Maxwellian core, a suprathermal halo at all pitch angles, and the strahl --- a magnetic field-aligned beam covering the same energy range as the halo. In the absence of specific magnetic structures, such as switchbacks \citep[e.g.][]{bale19,kasper19}, the strahl is directed anti-sunward. Switchbacks are sudden magnetic deflection of the solar wind, which are ubiquitous in the young solar wind. 

The strahl is observed to broaden with increasing radial distance from the Sun \citep{hammond_variation_1996, graham_evolution_2017}, which goes against the conservation of the magnetic moment and suggests that wave-particle interactions are operating. Moreover, the relative density of the halo is increasing while the relative density of the strahl decreases \citep{maksimovic_radial_2005, stverak_radial_2009}. These observations suggest that the scattering of the strahl feeds the halo, although this process might not be the only source to explain the halo formation \citep{abraham_radial_2022}. The scattering of the strahl consequently regulates the solar wind heat flux that is mostly carried by this suprathermal population in the fast wind \citep{scime_regulation_1994}.

The whistler generation mechanisms in the solar wind are still debated; furthermore, the dominant plasma instability generating the whistlers may depend on the heliocentric distance. The observations of whistlers and their specific properties such as wave normal angle (WNA), propagation direction, amplitude, frequency, and occurrence, can reveal the type of instability involved. Whistler waves with small WNA, in other words with a $k$-vector quasi-parallel to the background magnetic field, seem to be the most frequently reported \citep{stansby_experimental_2016, 2019ApJ...878...41T, kretzschmar_whistler_2021} when detected in the magnetic field data. However, studies where whistlers are detected in the electric field \citep[e.g.][]{breneman_observations_2010, cattell_narrowband_2020-1} show a large proportion of oblique waves since these whistlers are nearly-electrostatic. Quasi-parallel whistlers can be produced by the whistler heat-flux instability (WHFI) \citep{gary_heat_1975, gary_whistler_1994, feldman_electron_1976, roberg-clark_scattering_2019, roberg-clark_wave_2018}. This instability is inherently favored by the expansion of the solar wind \citep{micera_role_2021} and is generated by the counter-streaming electrons of the core and halo. Quasi-parallel whistlers can also be created by a resonant instability driven by the relaxation of the sunward deficit in the core eVDF \citep{bercic_whistler_2021} when the WHFI cannot be triggered \citep{halekas_sunward_2021}. However, while the instability driven by the sunward electron deficit can lead to a decrease of the total heat flux \citep{bercic_whistler_2021}, it is unclear whether it can contribute to the scattering of the strahl.

While quasi-parallel whistlers appear to dominate in the young solar wind, there is also evidence of oblique whistlers \citep{agapitov20, cattell_narrowband_2021} that, contrary to parallel whistlers, do not need to propagate counter-propagate with the strahl in order to interact with it. One candidate for the generation of these waves is the oblique whistler instability, or fan instability, that is generated by anomalous cyclotron resonances of electrons \citep{vasko_whistler_2019} and can significantly scatter the strahl. However, the existence of the fan instability in the solar wind is still under debate \citep{jeong_stability_2022}. There exist other potential mechanisms for oblique whistler generation; \citet{micera_particle--cell_2020} demonstrated that the WHFI can also create short-lived oblique whistlers that will scatter the strahl. Oblique generation apart, they also showed that sunward quasi-parallel whistlers can originate from the relaxation of the oblique whistlers. \citet{sauer_beam-excited_2010} also proposed an alternative mechanism for the generation of oblique whistlers in the presence of electron beams that propagate with velocities greater than twice the Alfvén velocity.

On top of the WNA, the direction of propagation is a crucial parameter for evaluating the efficiency of whistler waves in scattering the strahl. Enhanced pitch-angle scattering has been observed in the presence of whistler waves \citep{pagel_scattering_2007, cattell_parker_2021, jagarlamudi_whistler_2021}. Reports of whistler waves at heliocentric distances greater than 50 solar radii overwhelmingly show the waves dominantly propagate anti-sunward \citep[e.g.][]{lacombe_whistler_2014, 2019ApJ...878...41T, kretzschmar_whistler_2021}. However, the co-propagation of the anti-sunward waves with strahl electrons makes them inefficient at scattering the strahl \citep{verscharen_self-induced_2019}. Recent observations by PSP at 35 solar radii indicate that there is a population of sunward propagating whistlers in the young solar wind \citep[]{agapitov20, cattell_parker_2021, colomban_2022}; the study by \citet{agapitov20} reports on the collocation of oblique sunward propagating waves with local magnetic dips, which suggests the possibility of their local generation from the temperature anisotropy of a trapped hot electron population. These waves, even with a lower occurrence than anti-sunward waves, could significantly contribute to strahl scattering. Further studies are needed to estimate their occurrence rate and wave 
parameters for a proper evaluation of their scattering efficiency in the solar wind.

Solar wind observations combined with numerical simulations have also shown that the wave packet structure (spectrum of amplitudes) has  an effect on the scattering of the strahl \citep{saito_all_2007, saito_whistler_2007}. The recent simulations studies of \citet{cattell_modeling_2021, vo_stochastic_2022}, have further
showed the effect of a spectrum of $k$-vectors.

In the present paper, we aim to study the properties of whistler waves in the young solar wind; we use the continuous-time and large frequency coverage offered by the DC cross-spectral matrices based on the measurements of the Search-Coil Magnetometer \citep[SCM,][]{jannet_measurement_2021} analyzed by the Digital Field Board \citep[DFB,][]{malaspina_digital_2016} of the FIELDS \citep{Bale2016} experiment on board Parker Solar Probe. We cross-validate our results by comparing different data products as explained in Section~\ref{sec:method}. In Section~\ref{sec:results}, we present the detailed analysis of three whistler events with variable characteristics. This allows us to demonstrate the estimation of the WNA from the cross spectra by comparing our results with time-series of waveform measurements. We can thus confidently present statistics on the WNA of the whistlers and their frequent collocation with magnetic dips in Section~\ref{sec:stats}.
Our study is thus complementary of other statistical studies analysing the same dataset such as \citet{jagarlamudi_whistler_2021} and \citet{cattell_parker_2022} that had a different focus.
Finally, we discuss the implications in terms of wave-particles interactions Section~\ref{sec:discussion} and summarize our results in Section~\ref{sec:conclusion}.

\section{Data description and analyses techniques}\label{sec:method}
\subsection{Parker Solar Probe measurements}

The FIELDS suite onboard Parker Solar Probe \citep[]{Bale2016, malaspina_digital_2016, pulupa17} carries a series of instruments able to measure the electric and magnetic fields from DC up to 20 MHz \citep[]{Bale2016}: the Search-Coil Magnetometer (SCM), the Electric Field (EF) antennas, and the Fluxgate MAGnetometer (MAG). The SCM is a three orthogonal axes magnetometer measuring the fluctuations of the magnetic field between \SI{3}{Hz} and \SI{1}{MHz}. An in-depth collection of the major SCM first results and instrument description is available in \citet{dudok_de_wit_first_2022}. The data produced by the DFB from SCM and EF measurements are waveforms with a 292.97 s$^{-1}$ sampling rate, 3.5-second burst intervals with 150~000 s$^{-1}$ sampling rate, and spectral data (amplitude spectrum of electric field and spectral matrices of the magnetic field).

Here, we focus on properties of whistler waves as expressed in the magnetic DC cross-spectral matrices. The cross spectra contain the real and imaginary parts of the six Fourier cross products and the three auto spectra for the different spatial axes. From these we can retrieve the full spectral matrices for detecting whistler wave packets and derive their polarization properties. The cross-spectral products are computed on board from fast Fourier transform calculations averaged over 28-second bins (\SI{27.96}{s} cadence). They cover the frequency band \SI{23}{Hz} to \SI{4541}{kHz}, which embraces most of the whistler frequency range for the solar wind conditions that are explored in the present analysis ($f_{ce}<\SI{3000}{Hz}$). We complete our analysis with other types of DFB data products derived from the SCM measurements.  Because of the relatively low cadence of the cross spectra compared to the characteristic timescales of the fluctuations in the solar wind background magnetic field, we compliment our cross-spectral analysis with Band-Pass Filtered (BPF) measurements that offer a cadence of \SI{0.87}{s}. BPF measurements provide the amplitude of the wave magnetic field in specific spectral bands for one SCM component $\mathrm{B_u}$ (in the sensor frame). These allow us to precisely locate the wave packets within the 28-seconds cross-spectral bins, which is essential to compute the WNA (see \ref{sec:method_theta}). 
We do not present statistics on whistlers as detected in the BPF. This was already done in e.g. \citet{jagarlamudi_whistler_2021}, in particular on the wave amplitude and duration. However, the BPF-derived amplitudes presented in Section~\ref{sec:wave_normal_angle} allow us to validate the consistency of our results with other studies.
Finally, we use waveforms of the magnetic field for the validation of the cross-spectral analyses and the electric field to determine the direction of propagation of the wave packets.

The survey waveforms are continuously available at increasing sampling frequency near perihelion from \SI{73}{Hz} to \SI{292.97}{Hz}. This limits significantly the number of whistler waves detected in the cross-spectral data that we can compare with continuous waveforms. However, we benefit from the availability of high frequency burst waveforms. Several tens of bursts at \SI{150}{kHz} are available per day. These bursts do not capture all the whistlers waves, they are down-selected in the FIELDS memory, in order to always keep the bursts of highest quality \citep{Bale2016, malaspina_digital_2016}. The waveforms presented in the paper (MAG, SCM, and EF) are shown in the RTN frame: $R$ is radial and points away from the Sun, the tangential $T$ component is the cross-product of the solar rotation vector with $R$, and the normal $N$ component completes the right-handed set and points in the same direction as the solar rotation vector. 

We also use the solar wind background vector magnetic field from the MAG instrument; the electron density derived from the Radio Frequency Spectrometer \citep[RFS,][]{pulupa17} measurements with the Quasi-Thermal Noise (QTN) technique \citep{moncuquet20}, at a cadence of about 7 seconds; the radial proton velocity from the Solar Wind Electrons Alphas and Protons (SWEAP) suite instrument \citep{Kasper2016} Solar Probe Cup \citep[SPC,][]{case_solar_2020}, at a cadence of \SI{0.87}{s}.

The electric field is measured by the electric fields instrument (EF) consisting of two pairs of dipole electric field antennas oriented in the $\mathrm{TN}$-plane and extending beyond the PSP heat shield, and a fifth antenna located behind the heat shield on the instrument boom; the location of antenna $5$ in the wake of PSP means the $\mathrm{R}$-component is susceptible to detrimental interference by the wake electric field and cannot be reliably interpreted \citep{Bale2016}. The $\mathrm{R}$ component of the wave electric field, not used in the present study, can be reconstructed from $\Vec{E}\cdot\Vec{B}=0$ in the whistler frequency range.

The dataset we explore corresponds to the first encounter of PSP with the Sun from November 1, 2018, to November 11, 2018. During this time frame both the cross spectra and BPF data products were available for studying magnetic field fluctuations. For this first approach, the radial distance to the Sun spanned between 35.7 and 54 solar radii. After Encounter 1, an anomaly appeared in one of the SCM antennas, leading to the impossibility to make full use of the spectral matrices \citep[see][for more details]{dudok_de_wit_first_2022} for polarization analysis. 

\begin{figure}
    \centering
	\includegraphics[width=\linewidth]{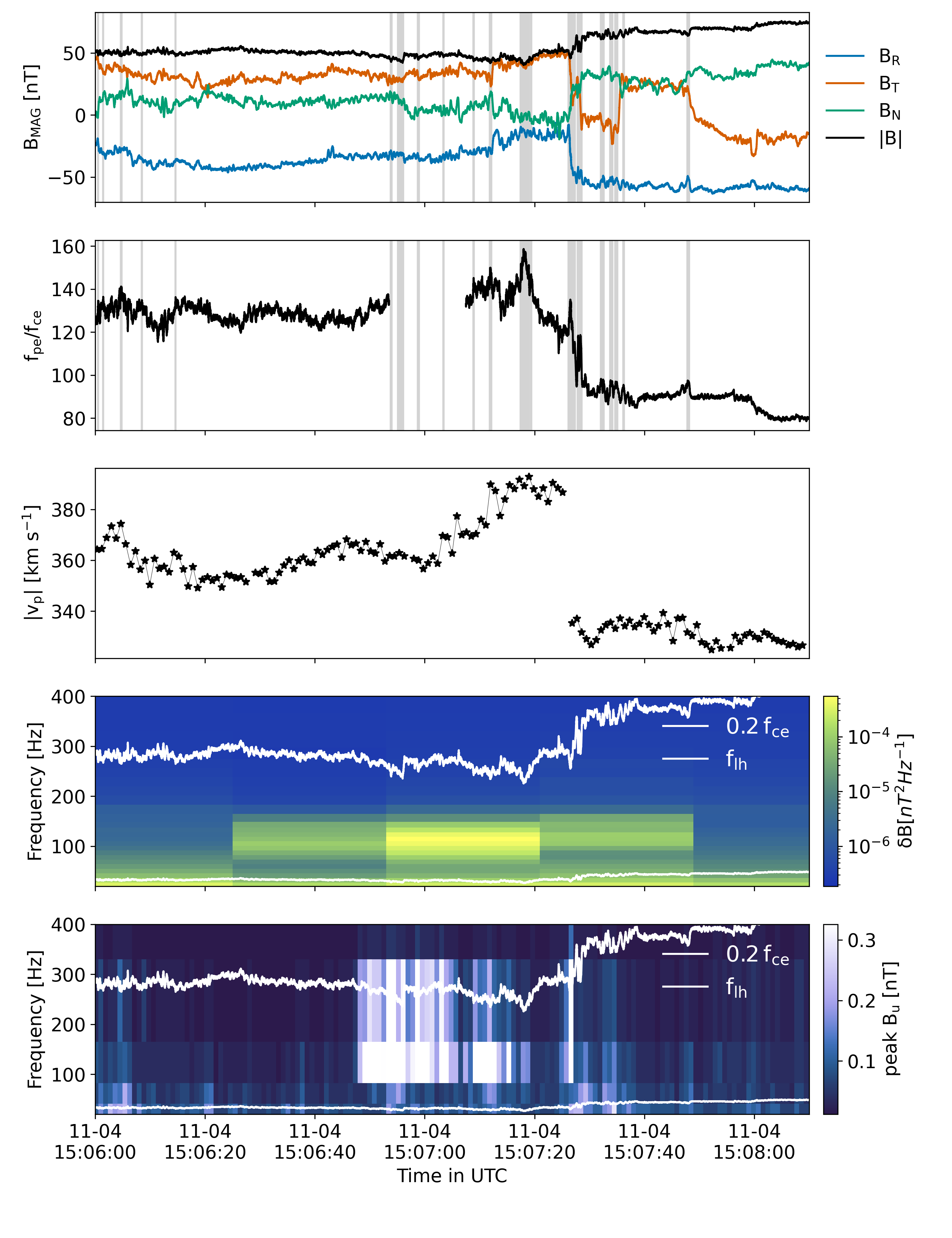} 
	\caption{Event~1: whistler waves observed on November 4, 2018, around 15:07 UTC. \textit{First panel:} solar wind background magnetic field components in the RTN frame and magnitude from the MAG instrument. \textit{Second panel:} $f_{pe}/f_{ce}$ ratio. On these two first panels, the gray bars highlight the presence of dips in the background magnetic field (see section~\ref{sec:method_dips}). \textit{Third panel:} magnitude of the proton velocity from the SWEAP/SPC instrument. \textit{Fourth panel:} trace of the cross-spectral matrix. \textit{Fifth panel:} peak value of the corresponding band-pass filtered measurements for the unique direction available for this data product. On the last two panels, the two white lines indicate 0.2 $f_{ce}$ (the local electron-cyclotron frequency)  and the lower-hybrid frequency $f_{lh}$, respectively.}
	\label{fig:event1_context}
\end{figure}

\begin{figure*}
    \centering
	\includegraphics[width=\linewidth]{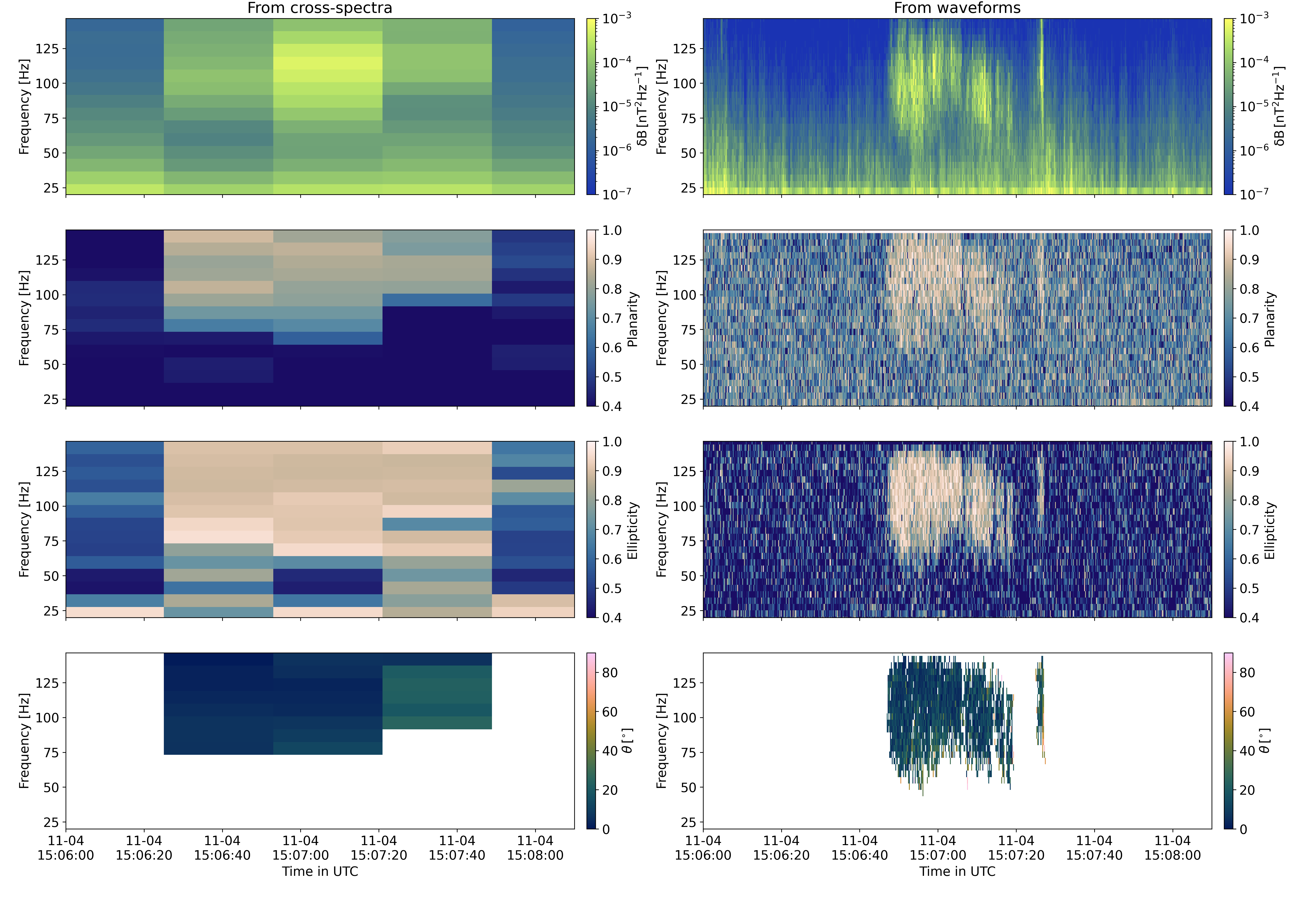} 
	\caption{Comparison of the polarization properties of event~1 derived from the cross-spectral measurements (left panels) and waveforms (right panels). The same time window was considered for the waveforms and spectra. \textit{First row:} trace of the cross-spectral matrix. \textit{Second row:} planarity. \textit{Third row:} ellipticity. \textit{Fourth row:} $\theta$ angle between the $k$-vector and the solar wind magnetic field. We only display the blocks of contiguous bins in frequency and time for which the ellipticity and planarity are higher than 0.6 and PSD at least 4 times higher than the ambient fluctuation level.} 
	\label{fig:event1_polarisation}
\end{figure*}

\subsection{Detection of whistler wave packets and polarization analysis} 
We detect whistler waves in the trace of cross-spectral matrices of the fluctuating magnetic field. We apply the following criteria to detect them:
\begin{itemize}
    \item First, we ensure that the Power Spectral Density (PSD) is at least four times above the ambient fluctuation level. The ambient level of magnetic field fluctuation is computed for each day of the dataset, by taking the median PSD, to account for the increasing power level of turbulence when approaching the Sun. This criterion was tested on the data and is effective to detect spectral bumps in the PSDs, that are characteristic of whistlers \citep[][]{jagarlamudi_whistler_2020};

    \item We then determine the ellipticity and planarity of the candidate waves to check if they are coherent waves and not enhanced turbulence. We use the Singular Value Decomposition (SVD) technique as described in \citet{santolik_singular_2003}. The planarity is given by: $1-\sqrt{w_3/w_1}$; the ellipticity by: ${w_2/w_1}$, where $w_1, w_2, w_3$ are the singular values in descending order. We chose a conservative approach using thresholds at 0.6 for both; 

    \item Finally, we only keep detections with a frequency higher than the local lower hybrid frequency ($f_{LH}$). We do not expect to exclude a significant amount of Doppler-shifted sunward whistlers. Even extreme cases in \citet{agapitov20} were observed above $f_{LH}$.

\end{itemize}

In our analysis of the spectra, we detect whistlers waves as wave packets, namely individual bins in time and frequency. This definition account for the fact that within the 28-second bin, several whistlers waves can occur. They can be at a similar frequency (same frequency bin) but not happening  at the same time (see event~2 in Section~\ref{sec:results}). Moreover, in that way we can capture the different WNA at each frequency of a true wave packet. 
We would thus would like to emphasize that the number itself of wave packets presented in Section~\ref{sec:stats} should not be directly compared with other studies \citep[e.g.][]{cattell_parker_2022}.

\subsection{Determination of the wave normal angle (WNA)}\label{sec:method_theta}
The wave normal angle $\theta$ (WNA) - the angle between the wave normal (the $k$-vector) and the local background magnetic field $\Vec{B}$ of the solar wind is computed as follows: $\arccos{\left(\frac{|\Vec{k} \cdot \Vec{B}|}{||\Vec{k}||||\Vec{B}||}\right)}$. The $k$-vector is given by the minor axis direction derived from the SVD. Since we focus here on magnetic field measurements only, we do not get the absolute orientation of the wave propagation. $\theta$ thus lies between 0\deg\,and 90\deg. We will determine the absolute orientation of the wave propagation for selected cases by using burst waveforms in Section~\ref{sec:detailed_cases}. The cross spectra were computed onboard in a modified sensor frame, so we rotate the derived $k$-vectors to the MAG frame (see Appendix~\ref{sec:appendix}). 

Whistler waves in the young solar wind are found to be intermittent \citep{jagarlamudi_whistler_2021} and to sometimes to occur simultaneously with magnetic field dips and/or boundaries of magnetic deflections such as switchbacks \citep{agapitov20}.
We thus have to choose a representative vector of the background magnetic field within each 28-second cross-spectral bins. To achieve this, we weight the average of the magnetic field vectors over the 28-second windows by the PSD computed from the peak value of BPF measurements (only when the corresponding PSD in the BPF is above 70 of the daily median PSD from BPF). The background magnetic field chosen for each cross-spectral bin then corresponds to the background magnetic field at the time of occurrence of the whistler waves.

This method was already used in \citet{froment_direct_2021} and \citet{dudok_de_wit_first_2022}. In the present paper, we detail the validation of this method by comparing the WNA obtain with cross spectra with those obtained with the waveform for a few examples (Section~\ref{sec:detailed_cases}).

We categorize whistlers as oblique when their  WNA $\theta > 45\deg$.
This threshold is the same for the entire dataset even though the Gendrin angle $cos \, \theta_G=2f/f_{ce}$ \citep{gendrin_guidage_1961} varies from 45\deg to 87\deg (for $f/f_{ce} = 0.35$ and $f/f_{ce} = 0.03$, respectively, which are the maximum and minimum $f/f_{ce}$ ratios encountered in our analysis).

\subsection{Collocation with magnetic dips}\label{sec:method_dips}
We search for possible collocations with magnetic dips for the detected whistler waves. We choose to implement a rather simple detection technique since these detections will exclusively serve to highlight the presence of magnetic dips in our whistler statistics. We look for a significant decrease of $|\mathrm{B}|$ compared to the ambient fluctuation level in the magnetic field from MAG. We first apply a low-pass filter on the background solar wind magnetic field magnitude $|\mathrm{B}|$. We define the deviation of $|\mathrm{B}|$ from this low pass-filtered version of $|\mathrm{B}|$ as: $\mathrm{(|B|-|B|_{filt})/|B|_{filt}}$. We then locate where this relative depth drops below $-0.05$. With this method, we miss shallow magnetic holes, even though we detect the smaller ones that are usually embedded in these larger-scale magnetic holes (see Section~\ref{sec:magnetic_dips}). 


\section{Validation of the processing technique}\label{sec:results}

\subsection{Detailed analysis of events}\label{sec:detailed_cases}
We present in this section the detailed analysis of three cases of whistler waves based on the DFB cross spectra and for which we compare the polarization processing results with the ones derived from waveform measurements. This comparison allows us to validate the use of cross spectra with a 28-second time resolution to compute the WNA of transient whistlers with the method described in Section~\ref{sec:method_theta}. 

These events are representative of the diversity of the whistlers detected in the datasets explored in terms of obliquity (quasi-parallel/oblique waves), intermittency, frequency and collocation with magnetic dips. 

\subsubsection{Event 1: quasi-parallel whistlers collocated with magnetic dips at the boundary of a switchback}

The first event we analyze occurred on November 4, 2018, around 15:07~UTC.
Figure~\ref{fig:event1_context} presents the background solar wind context and the DFB spectra for this whistler wave packet. The trace of the cross-spectral matrix shows significant spectral power (on average 125 times above the ambient fluctuation level) 
for three consecutive 28-second bins, i.e. \SI{84}{s} in total. The magnetic field signature in the BPF measurements is almost continuous for about 30~s.
We also observe a transient wave packet that lasts about \SI{1.6}{s} and is collocated with a magnetic dip (a 16\% drop of the local magnetic field magnitude). There are several magnetic dips detected with our method for this event. We notice that some also coincide with a local increase of $f_{pe}/f_{ce} \sim \mathrm{n_e}^{1/2}/\mathrm{B}$, however, this is only due to the magnetic decrease. $f_{pe}/f_{ce}$ is important for scattering effects \citep[e.g.][section 4.]{2016SSRv..200..261A}. The cadence of the density measurement is not sufficient to show a local density increase. Event~1 is located near the trailing edge of a switchback, characterized by a deflection in the $\mathrm{B_R}$ component. Moreover, we notice that the proton velocity that is on average about \SI{360}{\kms} goes up to about \SI{390}{\kms} right before this boundary, before decreasing to about \SI{330}{\kms}. The $f_{pe}/f_{ce}$ ratio is also decreasing (here both due to a decrease in density and increase of $|\mathrm{B}|$). The frequency of the whistlers detected in the cross spectra is ranging from \SIrange{73}{169}{Hz} (0.06-0.13~$f_{ce}$) in the spacecraft frame. This means that most of the whistlers for this particular event can also be studied by using survey waveforms. Their sampling rate was \SI{292.97}{Hz} near perihelion, we thus have access to wavepackets with frequencies below the Nyquist frequency that is \SI{146}{Hz} in the spacecraft frame. No burst waveforms were recorded during this interval under study. We construct cross spectra from these waveform measurements by using a short-time Fourier transform (STFT). There is no overlap of the segments of the spectrogram, but to increase the signal-to-noise ratio we average two consecutive spectra. After processing, the duration of each segment is \SI{0.2}{s}.

In the spectrogram constructed from the waveforms, we observe two main wave packets, as was already revealed in the BPF data. The first one has a duration of about \SI{34}{s} and another one of about \SI{2}{s}. Figure~\ref{fig:event1_polarisation} shows the result of a polarization analysis from the cross spectra on the left panels and waveforms on the right panels. On the bottom panels, we show the WNA when the planarity and ellipticity are higher than 0.6 and when the PSD is at least 4 times above the ambient fluctuation level. The median values of planarity and ellipticity are above 0.8. According to our selection criteria, these waves are quasi-parallel whistlers. From the cross spectra, the WNA goes up to 24.7\deg, with a median value of 6.2\deg. From the waveform, the median value is higher at 15.7\deg, and a few percent of the whistlers are oblique (2.7\%)

\begin{figure}
    \centering
	\includegraphics[width=\linewidth]{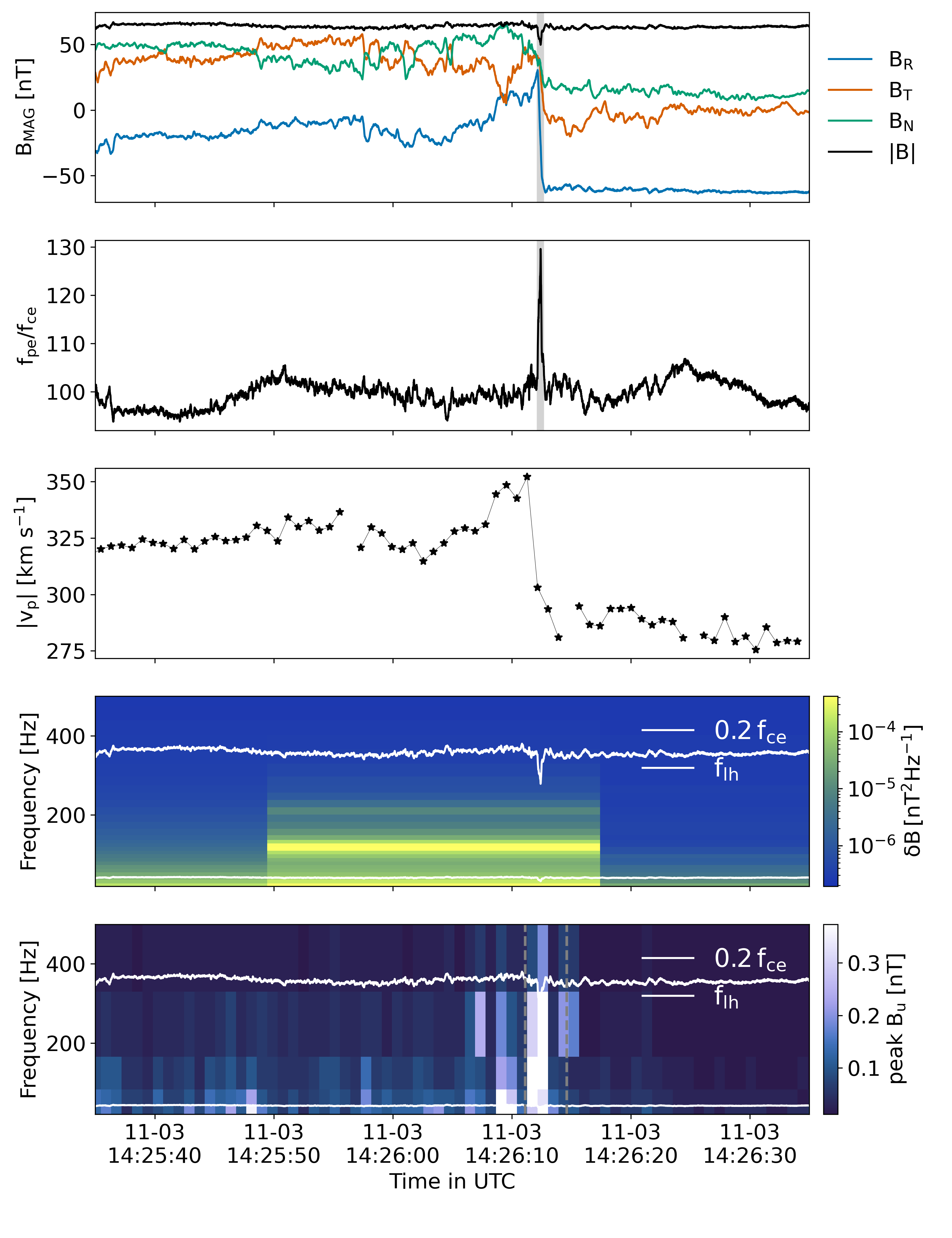} 
	\caption{Event 2: whistler waves observed on November 3, 2018, around 14:26 UTC. The different panels are the same as in Figure~\ref{fig:event1_context} for event~1. The two gray dashed lines delimit the burst window presented in Figure~\ref{fig:event2_polarisation}.}
	\label{fig:event2_context}
\end{figure}

\begin{figure*}
    \centering
	\includegraphics[width=\linewidth]{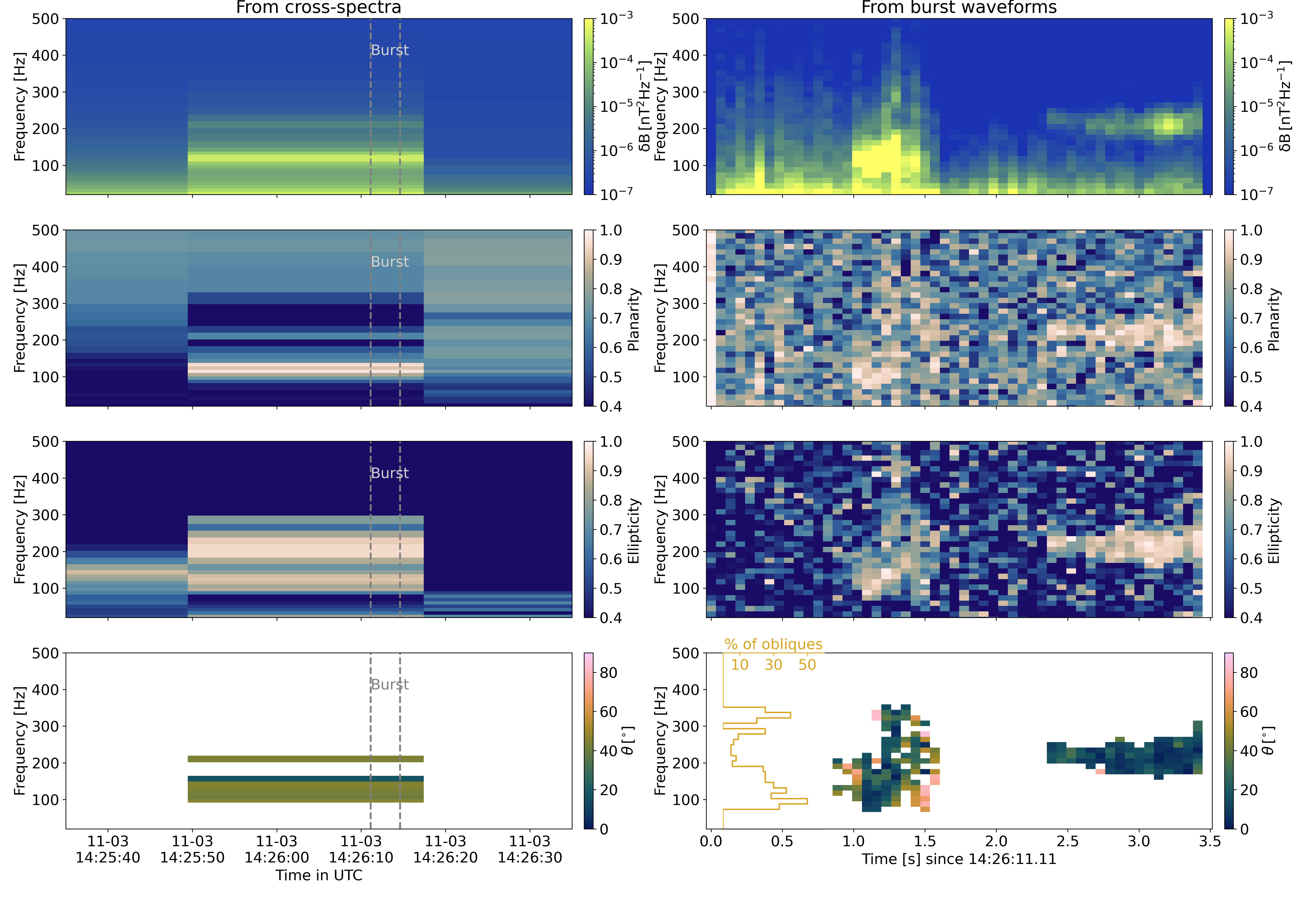} 
	\caption{Polarization properties derived from the magnetic field measurements using the SVD technique for event~2. The different panels are the same as in Figure~\ref{fig:event1_polarisation} for event~1. However, here the time window considered in the right panels corresponds to the burst time window indicated in the left panels in between the two gray dashed lines. The percentage of oblique whistlers per frequency bin is displayed for the burst waveform analysis.}
	\label{fig:event2_polarisation}
\end{figure*}

\begin{figure}
    \centering
	\includegraphics[width=\linewidth]{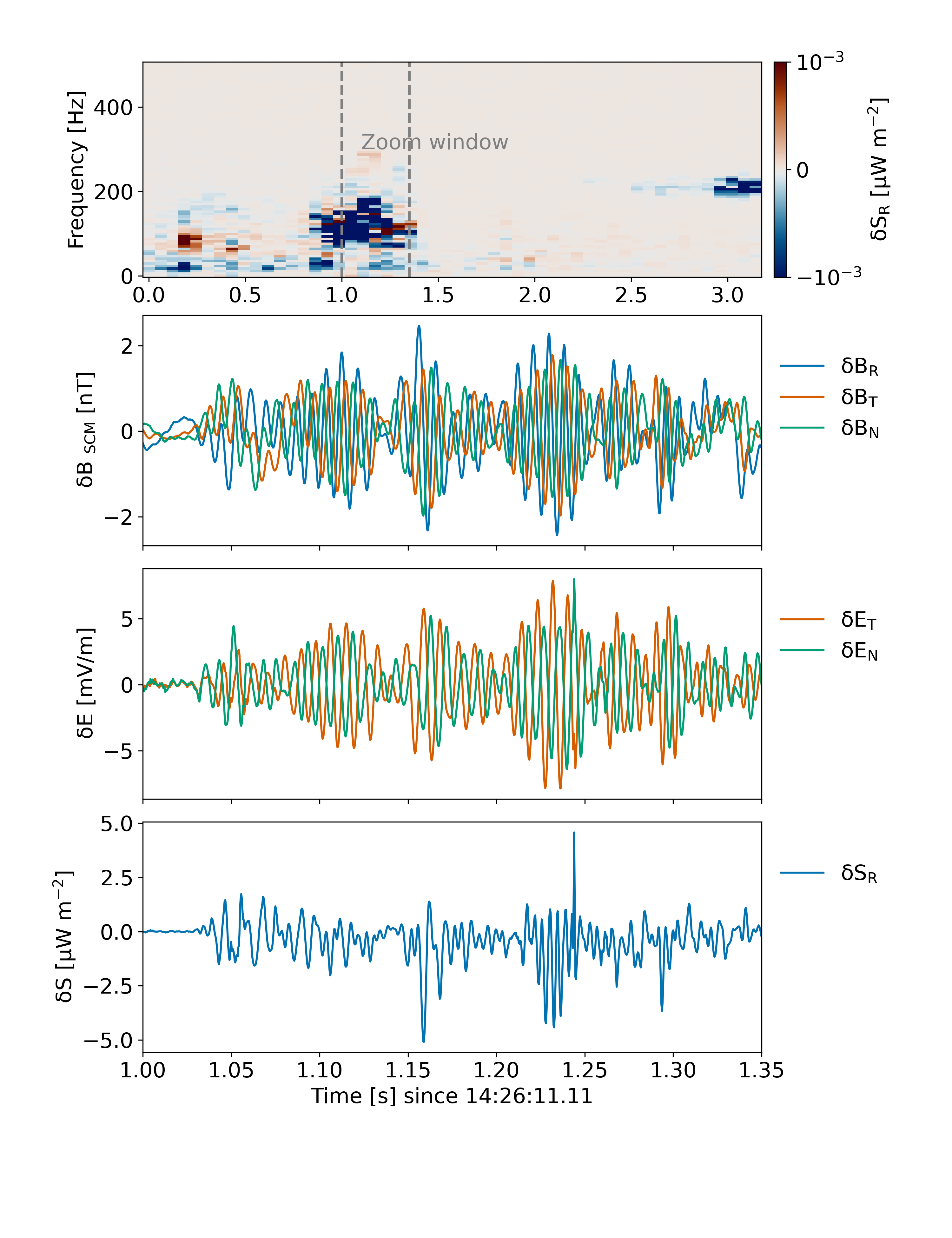} 
	\caption{Magnetic and electric waveforms, and Poynting flux for event~2. \textit{First panel:} Spectrogram of the R component of the Poynting Flux. The two gray dashed lines delimit the temporal window for which we display the waveforms. \textit{Second panel:} DFB magnetic burst waveforms from the SCM measurements. \textit{Third panel:} DFB electric burst waveforms from the EF measurements. \textit{Fourth row:} Waveform of the radial component of the Poynting Flux $\mathrm{S_R}$.}
	\label{fig:event2_waveforms}
\end{figure}

\subsubsection{Event~2: quasi-parallel and oblique whistlers collocated with magnetic dips at the boundary of a switchback} 

The second example we highlight here was recorded on November 3, 2018, around 14:26~UTC. The context of this detection is presented  in Figure~\ref{fig:event2_context}. The trace of the cross-spectral matrix shows an excess of spectral power for one 28-second bin (on average 186 times above the ambient fluctuation level). This is a strong signature even though the signature in the BPF data lasts only for about a few seconds by the end of the cross-spectral bin. The whistler wave packets are located at the trailing edge of a switchback (similar to event~1). These are also cotemporal with magnetic dips, with a local relative decrease of the magnetic field magnitude of 22\%. Magnetic dips are often observed at the boundaries of switchbacks \citep{agapitov20, froment_direct_2021}. A superposed epoch analysis on switchback events showed that the sharp switchback boundaries tend to produce a clear and distinct decrease in $|\mathrm{B}|$ at both the entry and exit of the switchback of $\sim 0.1 |\mathrm{B}|$ in average \citep[]{farrell_magnetic_2020}. This dips are naturally generated during switchback generation \citep[]{drake_switchbacks_2021} and propagation \citep[]{agapitov_flux_2022}. Similar to event~1, the proton velocity slightly goes up (from \SI{325}{\kms} to \SI{350}{\kms}) right before the switchback boundary and at the location of the whistlers. The $f_{pe}/f_{ce}$ ratio increases by about 30\% at the magnetic dip, both due to the local magnetic decrease and a local density increase. The frequency of the whistlers detected in the cross spectra is ranging from \SIrange{96}{237}{Hz} (0.05-0.14~$f_{ce}$) in the spacecraft frame. For this event burst waveforms are available.

The burst waveform interval starts at 14:26:11.1~UTC. The burst covers the main peaks seen in the BPF data. In particular, the wave packet in the magnetic dip as can be seen in Figure~\ref{fig:event2_context} (dashed interval on the BPF measurements). The polarization analysis results are presented in Figure~\ref{fig:event2_polarisation}. The spectrogram derived from the waveforms reveals two wave packets: the first one is cotemporal with the magnetic dip, and the second one is the start of the second wave packet seen in the BPF data during this interval. The planarity and ellipticity are quite high (above 0.7 and 0.8 respectively) for both the cross spectra and the waveforms. For the STFT performed on the burst waveforms, we choose a segment duration of \SI{68}{ms}. 

From the cross spectra, the whistler waves are found to be barely oblique, with a median WNA across the wave packets of 43.6\deg (maximum at 46.3\deg). From the burst waveforms, we obtain a mix of quasi-parallel and oblique whistlers. Even though the median value is lower for the waveform (18.2\deg) than for the cross spectra, we observe a significant amount of oblique whistlers (16\% over the two wave packets). While the second wave packet is quasi-parallel (median WNA 12\deg, with 1\% being oblique whistlers), in the first one collocated with the magnetic dip the WNA varies (median at 29.1\deg with 28\% being oblique whistlers). In Figure~\ref{fig:event2_polarisation} we display the percentage of obliques per frequency bins for both groups. It highlights that in the frequency band detected in the cross-spectra, about 30\% at least of the whistlers are oblique.
The WNA derived from the cross spectra can be understood as a snapshot of the more detailed distribution of WNA that we derive from the burst waveform. Since for this event, the whistler wave are located in a magnetic dip, our method (as explained in Section~\ref{sec:method_theta}) will capture a representative background magnetic field, for which the whistler waves are more intense, from which we compute the WNA. But we cannot capture the full length of the WNA inside the magnetic dip located at a switchback boundary as the vector direction changes quite dramatically. 

For this event, we can also use the electric field burst waveforms for computing the R-component of Poynting flux and thus determining the absolute direction of propagation of the waves. The magnetic and electric fluctuations as well as the R-component of Poynting flux $\mathrm{\delta S_R}$ for the first wave packet are presented in Figure~\ref{fig:event2_waveforms}. We also show the spectrogram of $\mathrm{\delta S_R}$ for the full burst duration. $\mathrm{\delta S_R}$ is significantly negative which means that these whistler waves are propagating sunward. We note that event~2 is thus very similar to the cases reported by \citet{agapitov20}, that is a collocation with a magnetic dip, sunward propagation, and a mix of quasi-parallel and oblique whistlers. However, since these waves are located at the boundary of a switchback, we note that sunward does not necessarily counter-propagating with the strahl \citep{colomban_2022}. Indeed, in the case of switchbacks, the strahl follows the orientation of the magnetic field during the deflection and depending on the configuration the strahl can become fully or partially sunward. Further investigation would be needed on this particular event in order to study the wave-particle interactions in details, which is outside of the scope of the present paper.

\begin{figure}
    \centering
	\includegraphics[width=\linewidth]{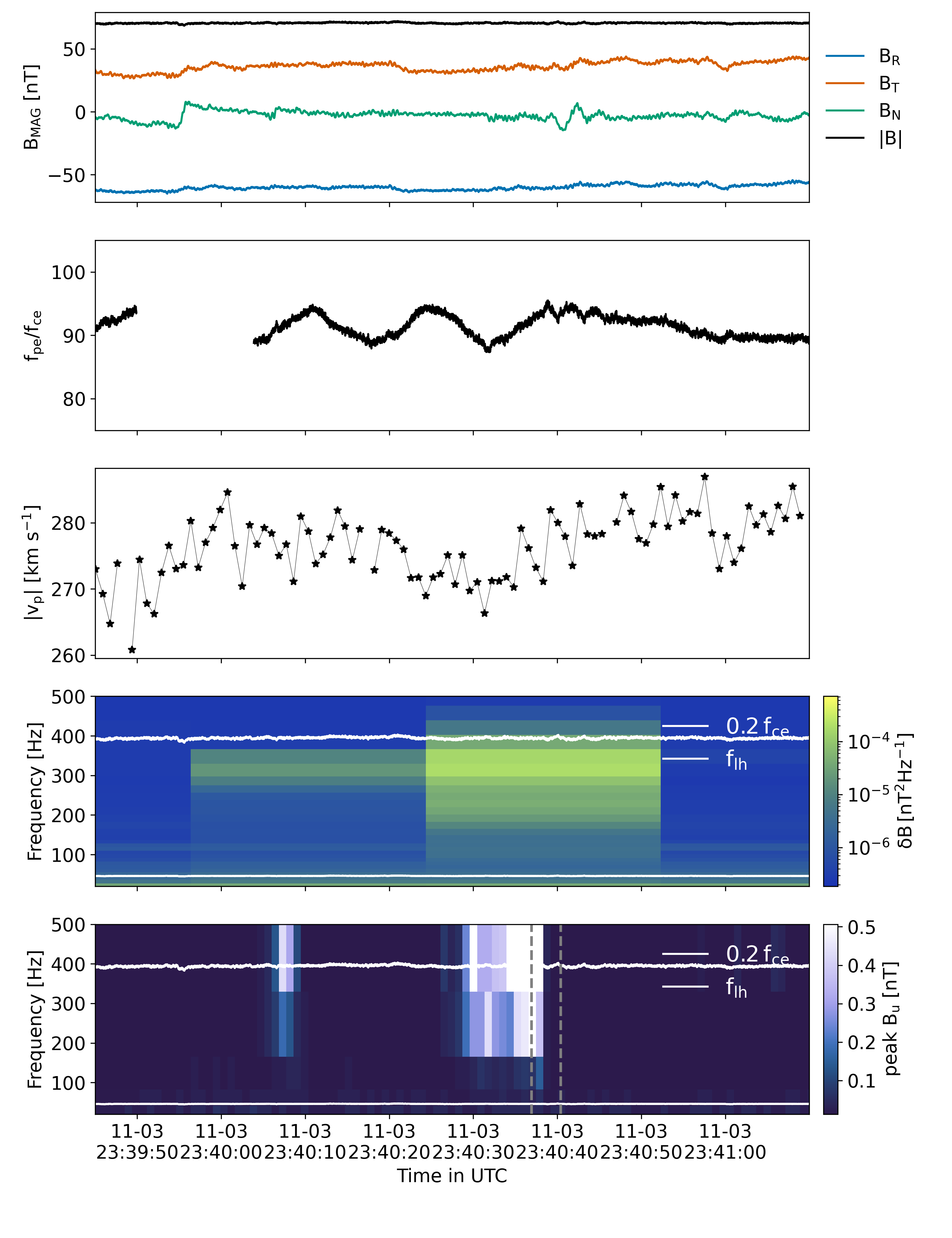}
	\caption{Event 3: whistler waves observed on November 3, 2018, around 13:50 UTC. The different panels are the same as in Figure~\ref{fig:event1_context} for event~1. The two gray dashed lines delimit the burst window presented in Figure~\ref{fig:event2_polarisation}.}
	\label{fig:event3_context}
\end{figure}

\begin{figure*}
    \centering
	\includegraphics[width=\linewidth]{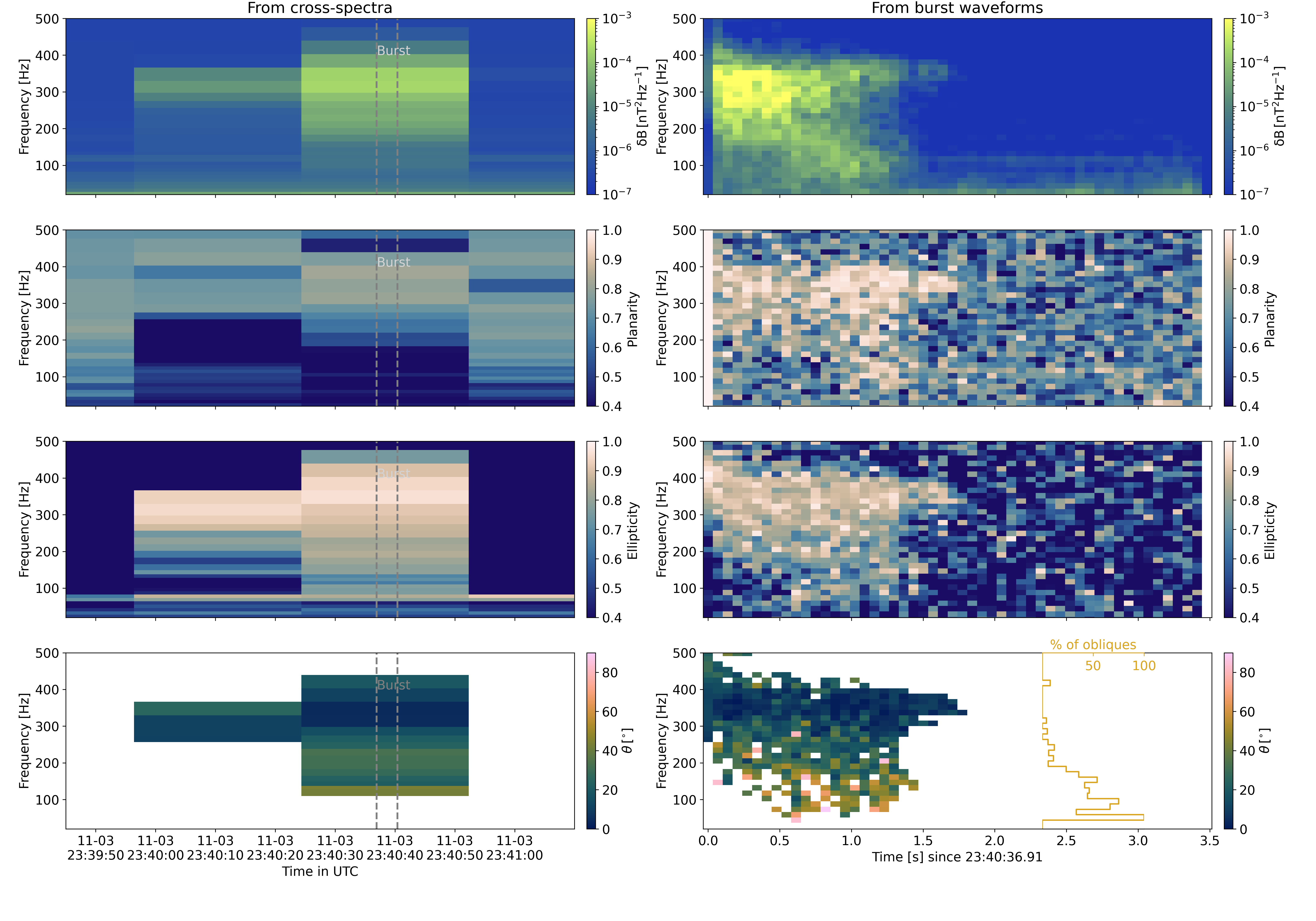} 
	\caption{Polarization properties derived from the SCM magnetic field for event~3. The different panels are the same as in Figure~\ref{fig:event1_polarisation} for event~1. However, here the time window considered in the right panels corresponds to the burst time window indicated in the left panels in between the two gray dashed lines. The percentage of oblique whistlers per frequency bin is displayed for the burst waveform analysis.}
	\label{fig:event3_polarisation}
\end{figure*}

\begin{figure}
    \centering
	\includegraphics[width=\linewidth]{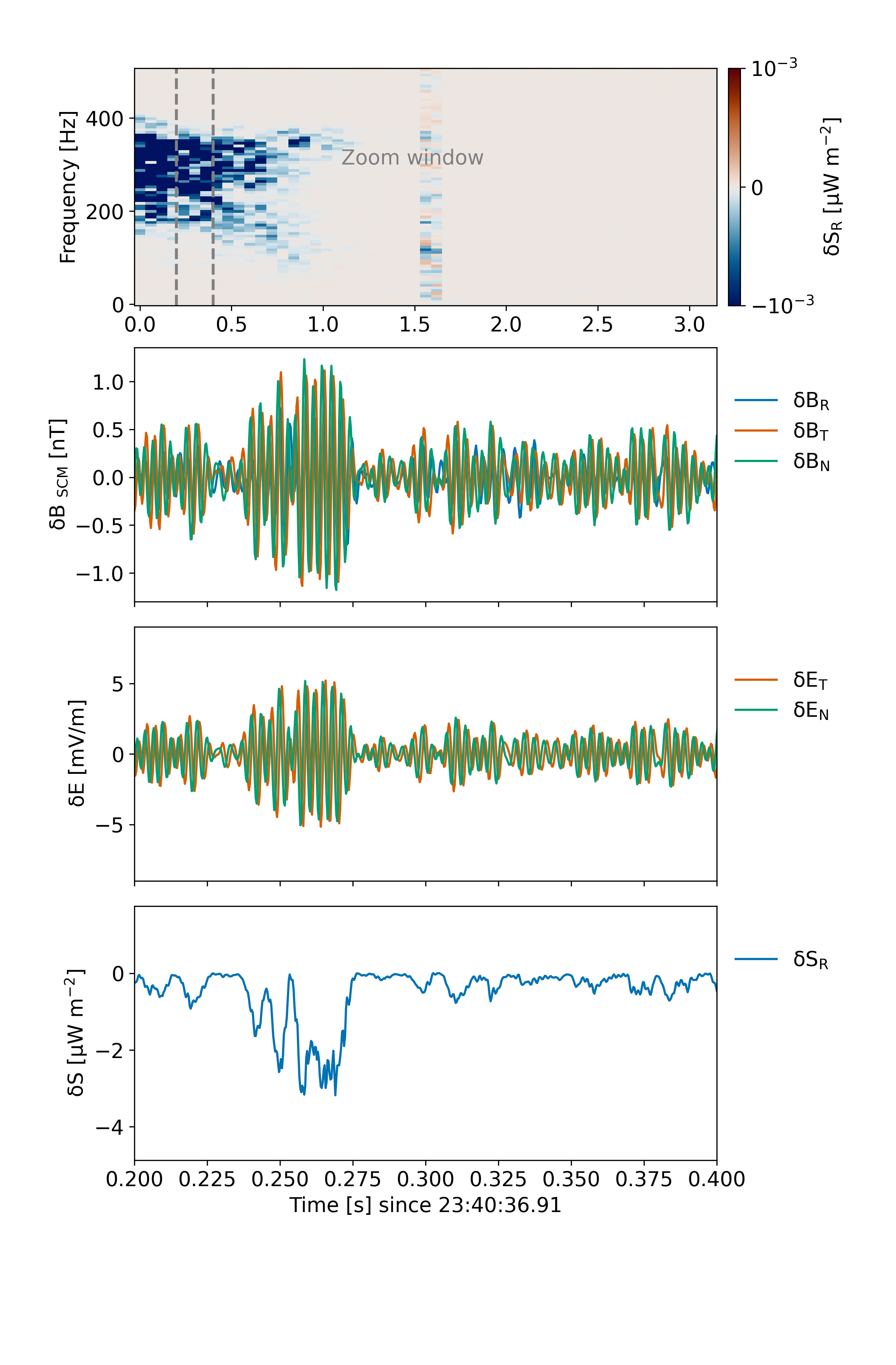} 
	\caption{Magnetic and electric field perturbation waveforms, and Poynting flux for event~3. The panels are the same as in Figure~\ref{fig:event2_waveforms} for event~2.}
	\label{fig:event3_waveforms}
\end{figure}

\subsubsection{Event 3: quasi-parallel and oblique whistlers}

The third event for which we detail the analysis here was recorded on November 3, 2018, around 23:40~UTC. Unlike the other two events, the whistlers detected for event~3 are associated neither with the boundary of a switchback nor with magnetic dips. As can be seen in Figure~\ref{fig:event3_context}, these whistler waves are encountered in a slower solar wind than for the two first examples. Indeed, the average wind speed was (\SI{276}{\kms}). This can be partially explained by the absence of the switchback, and its accompanying enhancement of velocity compared to the bulk solar wind, for the present example. The trace of the cross-spectral matrix shows an excess of power in two consecutive 28-second bins (on average 335 times above the ambient fluctuation level). In the BPF measurements, we observe a short whistler signature of about \SI{3.5}{s} corresponding to the first cross-spectral bin and a continuous burst lasting about \SI{9.6}{s} at the beginning of the second cross-spectral bin. There is no significant simultaneous variation of $f_{pe}/f_{ce}$. 
This event is also visible in the burst waveforms. As will be shown in Figure~\ref{fig:event3_polarisation}, the burst waveforms analysis leads to detections of whistlers at lower frequencies (down to \SI{40}{Hz}). We thus decided to relax the detection criteria by removing the threshold on the planarity for the cross spectra. The analysis in the cross spectra then covers the range \SIrange{114}{421}{Hz} (0.06-0.21~$f_{ce}$).

The burst waveform measurements start at 23:40:36.9~UTC. They cover the last part of the group of wave packets seen in the BPF measurements (see the dashed lines in Figure~\ref{fig:event3_context}). From the polarization analysis presented in Figure~\ref{fig:event3_polarisation}, we notice a wide-band wave packet that lasts for about \SI{1.8}{s}. Similar to the previous events, and due to our detection criteria, the ellipticity and planarity of the fluctuations is quite high from both the cross spectra and the waveforms (greater than 0.8). The median value of the planarity across the wave packet in the cross spectra is a bit lower, 0.6, but still significant. From the cross spectra, the waves are found to be mostly quasi-parallel with a median WNA of 22.6\deg. In the cross-spectra bin where we have burst waveforms, we see a gradual augmentation of $\theta$ from 3.9\deg\,to 43.8\deg. Such behavior is also very clear in the spectrogram constructed from the waveforms. Even though the median value of $\theta = 19.1\deg$ corresponds also to quasi-parallel whistlers, the percentage of obliques per frequency bins displayed on Figure~\ref{fig:event3_polarisation} clearly shows that more than 50\% of the whistlers are oblique below \SI{200}{Hz}. The whistlers seen in the burst waveforms seem to be divided into two groups: one high frequency packet above \SI{200}{Hz} that is quasi-parallel, and a second group at lower frequencies of \SIrange{40}{200}{Hz} that is oblique to highly oblique.

For this event, we also analyze the R-component of Poynting flux $S$ from the burst magnetic and electric field waveforms. The waveforms and spectrogram are presented in Figure~\ref{fig:event3_waveforms}. $\mathrm{\delta S_R}$ is significantly negative, which means these are sunward propagating whistler waves.
 
\begin{figure*}
    \centering
    \begin{tabular}{ccc}
    		\includegraphics[width=0.3\linewidth]{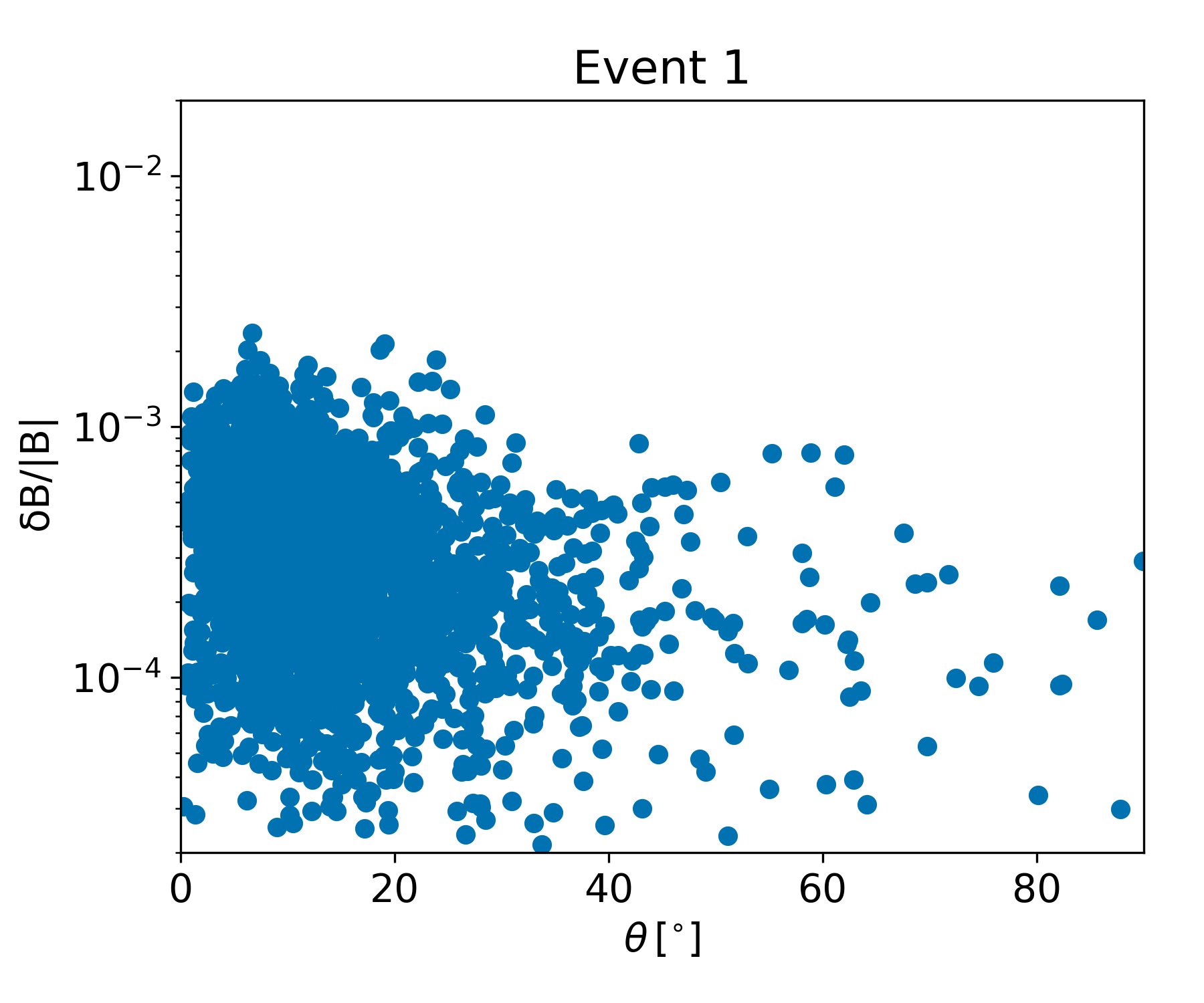}
     &  	\includegraphics[width=0.3\linewidth]{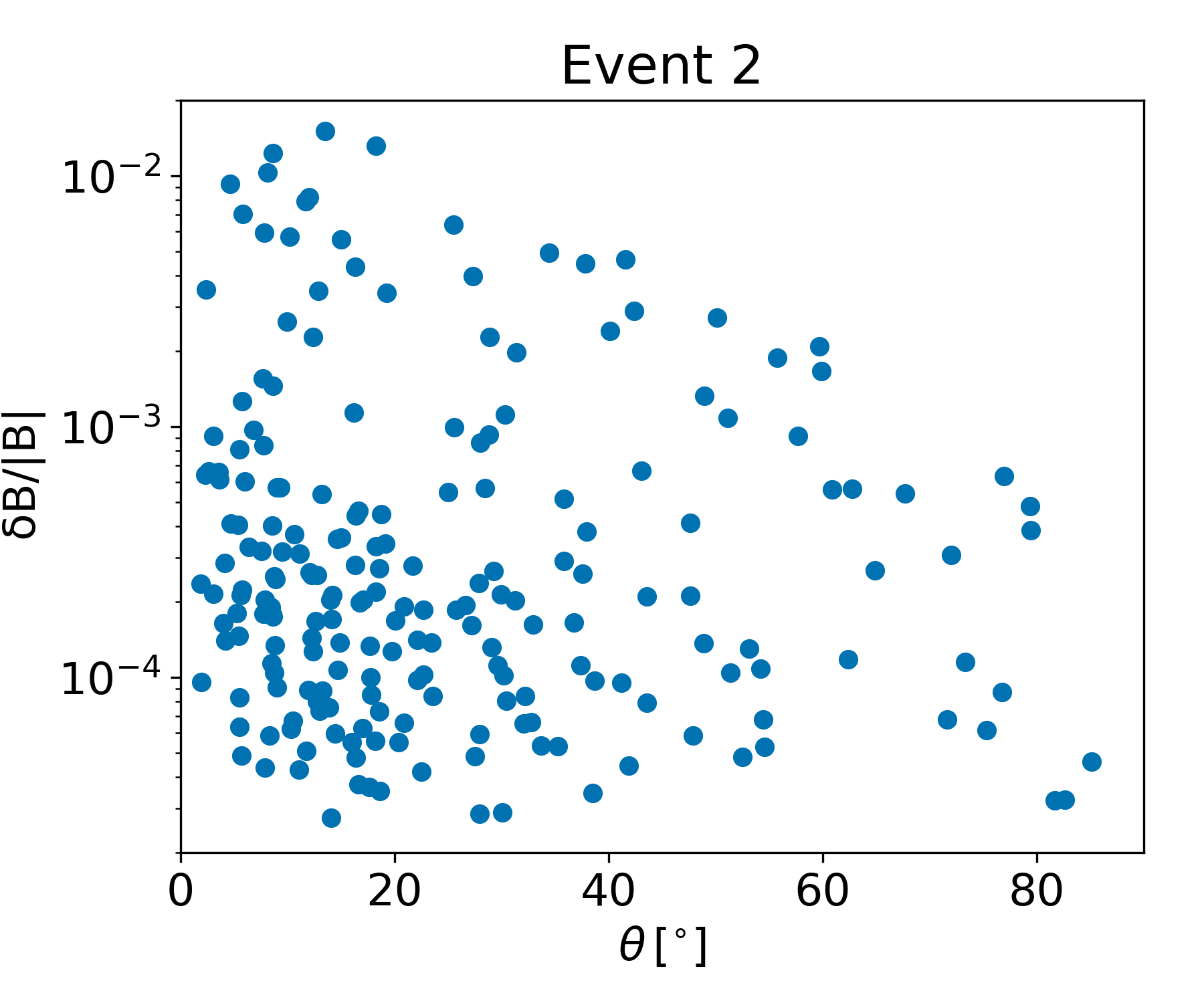}
     &  	\includegraphics[width=0.3\linewidth]{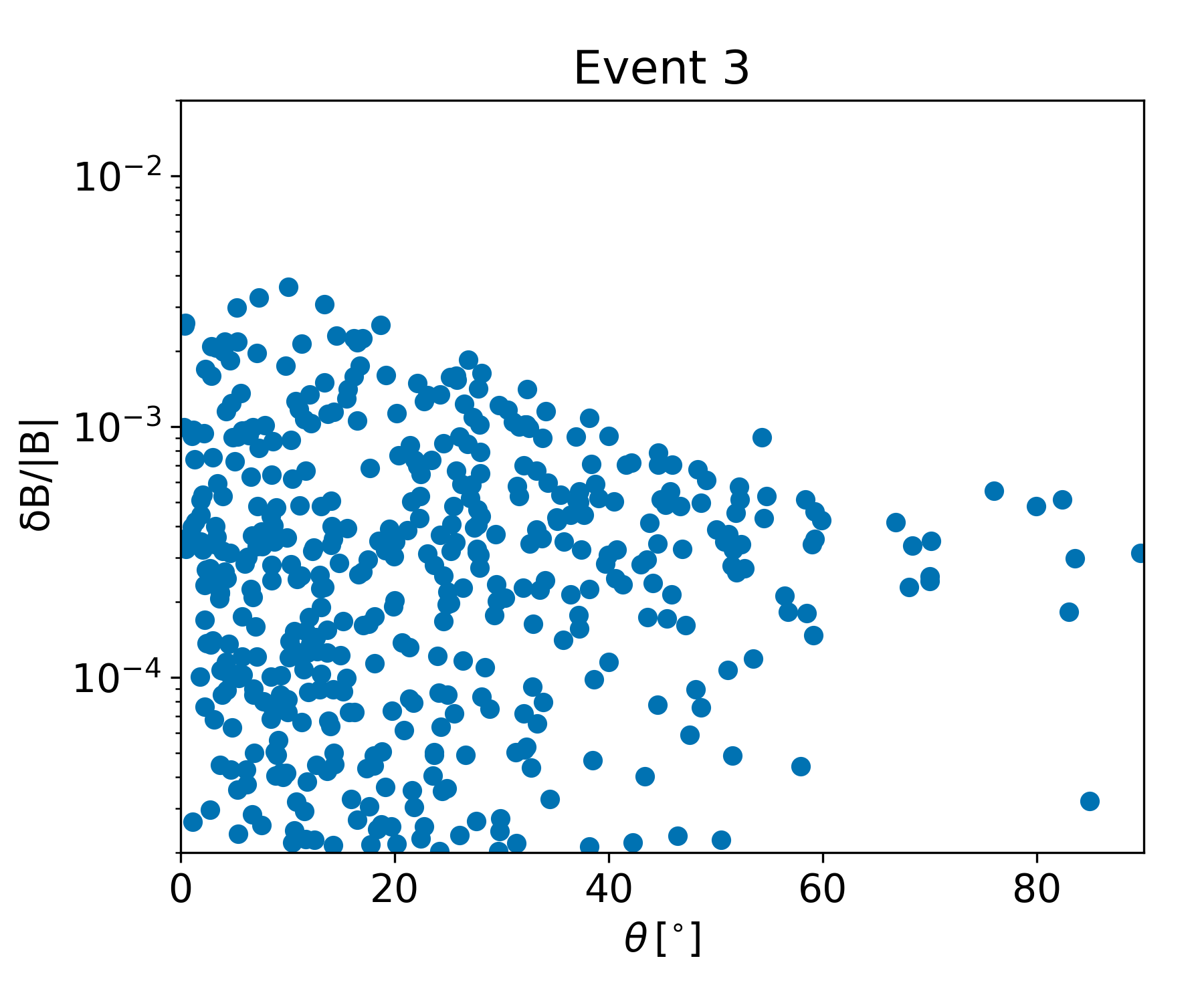}\\
    \end{tabular}
	\caption{Relative amplitudes of the magnetic fluctuations versus the WNA for the whistlers of the three events described in Section~\ref{sec:detailed_cases}. These waves correspond to the bins in the spectrograms constructed from the waveforms.}
	\label{fig:scatter_plots}
\end{figure*}

\subsection{Wave normal angle and amplitude of the fluctuations}\label{sec:scatter_amp_theta}

For each of the whistler bin in the spectrogram derived from the waveforms, we estimate the relative amplitude of the magnetic fluctuations as $\delta \mathrm{B}/|\mathrm{B}| = \sqrt{\mathrm{PSD \times \Delta f }}/|\mathrm{B}|$, where $|\mathrm{B}|$ is the magnitude of the background magnetic field from the MAG instrument. 

In Figure~\ref{fig:scatter_plots} we show the relative amplitude of the whistlers on the value of their WNA. First of all, we note that these amplitude values apply to individual waves bins in the spectrograms, which is different from amplitudes integrated over the entire frequency range as presented, for example, in \citet{2019ApJ...878...41T}, or directly from waveforms as in \citet{cattell_narrowband_2020-1}. These amplitudes are thus lower than those reported reported by others and should not be directly compared to them. 

For each event, the values of $\delta \mathrm{B}/|\mathrm{B}|$ are generally lower for oblique whistlers rather than for the quasi-parallel ones, up to about two orders of magnitude, which is expected \citep[]{verkh10, agapitov_statistics_2013}. Median values for the quasi-parallel and oblique populations of the three cases are quite close: $3 \times 10^{-4}$ and $2 \times 10^{-4}$, respectively for event~1, $2 \times 10^{-4}$  for both populations for event~2 and $3 \times 10^{-4}$ for both populations for event~3. However, event~2 is distinguished by its higher amplitudes of the fluctuations for both the quasi-parallel and oblique whistlers compared to the two other events. Indeed 9\% of the quasi-parallel whistlers have amplitudes above $4 \times 10^{-4}$, i.e. above the maximum values reached in event~1 and 3, and have a maximum of $2 \times 10^{-2}$. Also, 18\% of the oblique whistlers have amplitudes above $1 \times 10^{-3}$ with a maximum at $3 \times 10^{-3}$, a value only reached by the quasi-parallel whistlers for the other events.

In conclusion, the analysis of the amplitudes of fluctuations derived from the waveform analysis of the three events provides a further piece of evidence that the WNA determined from the cross spectra is consistent with the values determined from the waveform. This is in particular revealed by the large amplitudes of both quasi-parallel and oblique whistlers estimated for event~2, which was found to be barely oblique (the median value is 45.1\deg).

\section{Statistical properties}\label{sec:stats}

\subsection{General properties}\label{sec:global_stat}

In total, we detect 240 distinct whistler wave clusters. We define clusters as contiguous bins localized in time and frequency in the DFB cross spectra, similar to the examples presented in Section~\ref{sec:detailed_cases}. These represent in total 2710 individual wave packets. A few detections are removed after a visual examination of the spectra. These suspected spurious detections correspond to no clear peak that can be seen in the spectra and most likely correspond to enhanced levels of  turbulence. We finally get 2701 wave packets, which correspond to 232 whistler wave clusters. These waves have frequencies ranging from \SIrange{32}{531}{Hz} (median at \SI{141}{Hz}), that is, between 1.1~$f_{lh}$ and 0.35~$f_{ce}$. We note that 98\% of the whistlers have characteristic frequencies that are below 0.2~$f_{ce}$ which is consistent with previous studies \citep[]{jagarlamudi_whistler_2021, agapitov20, cattell_parker_2021, cattell_parker_2022}. Their frequency band extends between \SI{9}{Hz} and \SI{412}{Hz} (median \SI{82}{Hz}). The median planarity and ellipticity are 0.8 and 0.9, respectively. The wave clusters are observed most of the time (58\%) to have a duration within a single 28-second bin. The maximum duration of a wave cluster is 196 seconds. We note that such a long duration is very rare and can actually be misleading. \citet{jagarlamudi_whistler_2021} reported that the large majority whistlers waves (80~\%), as detected in the BPF measurements during PSP's encounter~1, last less than \SI{3}{s}, with a maximum duration of about \SI{70}{s}. Due to the low temporal resolution of the cross spectra during encounter 1, the duration and number of whistlers waves in the cross spectra are blurred: their number is underestimated while their duration is over estimated.

Whistlers are observed about 1.2\% of the time covered by the cross spectra. The whistler wave clusters are usually seen in groups lasting up to a few hours. After November 5, 2018, we notice a large decrease in the whistler occurrence rate: none appear for about 60 hours and only a few events are detected in the following 100 hours. This means that the large majority of whistlers in the cross spectra were detected during the inbound phase of the encounter. In the outbound phase of the encounter, the solar wind condition have changed: the averaged velocity has increased and several fast streams are encountered \citep{allen_solar_2020}. The lack of whistlers in this type of wind is consistent with the results of \citet{jagarlamudi_whistler_2020, jagarlamudi_whistler_2021} that showed that the occurrence of whistler waves is anti-correlated with the bulk solar wind velocity. 

Compared with waveform measurements, we note that we miss transient whistler wave bursts that are probably washed out in the cross-spectral averaged product. For example, the case described in Section~4 of \citet{dudok_de_wit_first_2022} that lasts for less than a second, does not appear in our detections.

\begin{figure}
    \includegraphics[width=\linewidth]{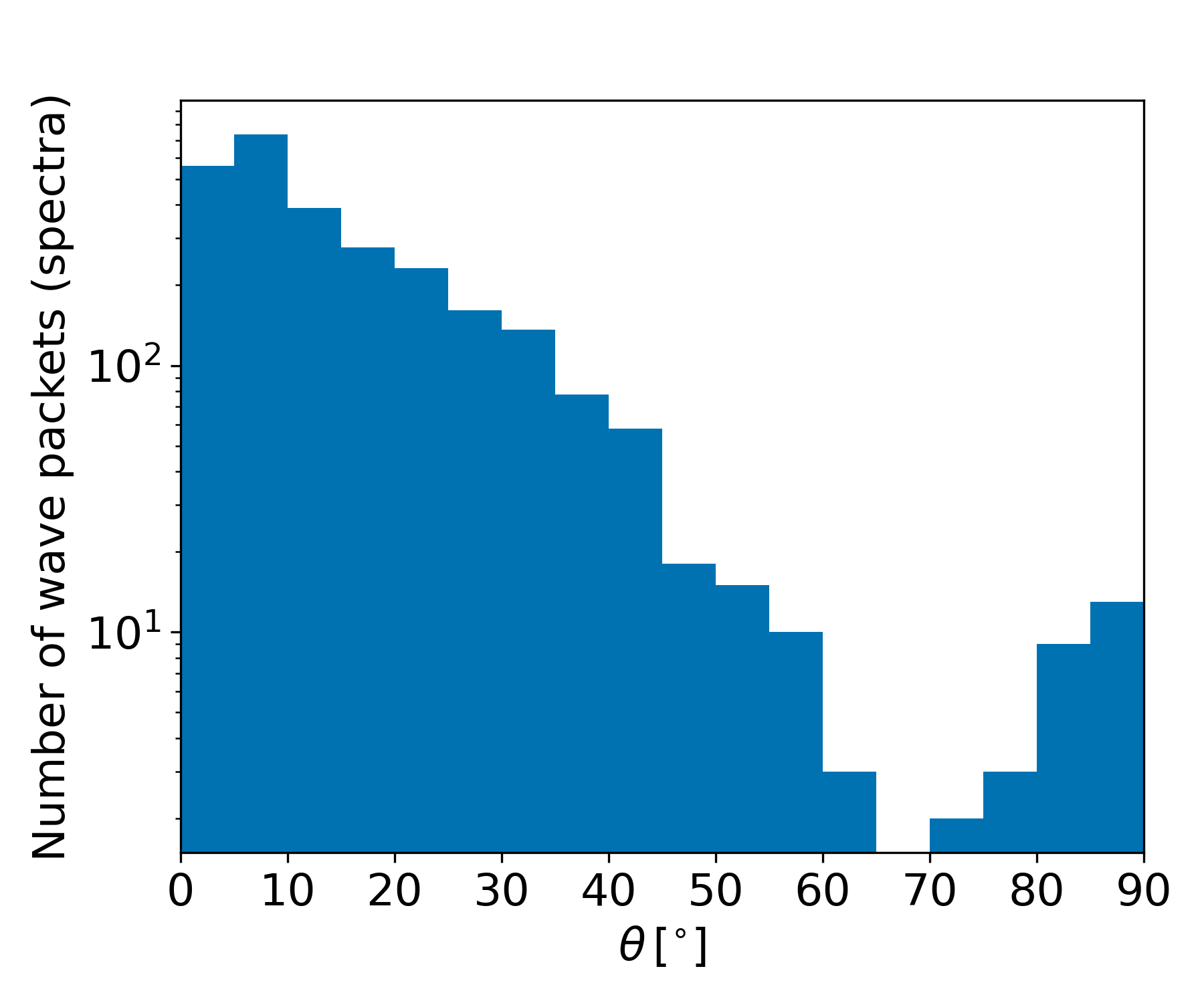} 
	\caption{Histogram of the WNA of whistler waves detected in the DFB cross spectra during the first encounter of PSP with the Sun (2701 wave packets in total).}
	\label{fig:stat_theta}
\end{figure}

\begin{figure*}
    \begin{tabular}{cc}
    	\includegraphics[width=0.45\linewidth]{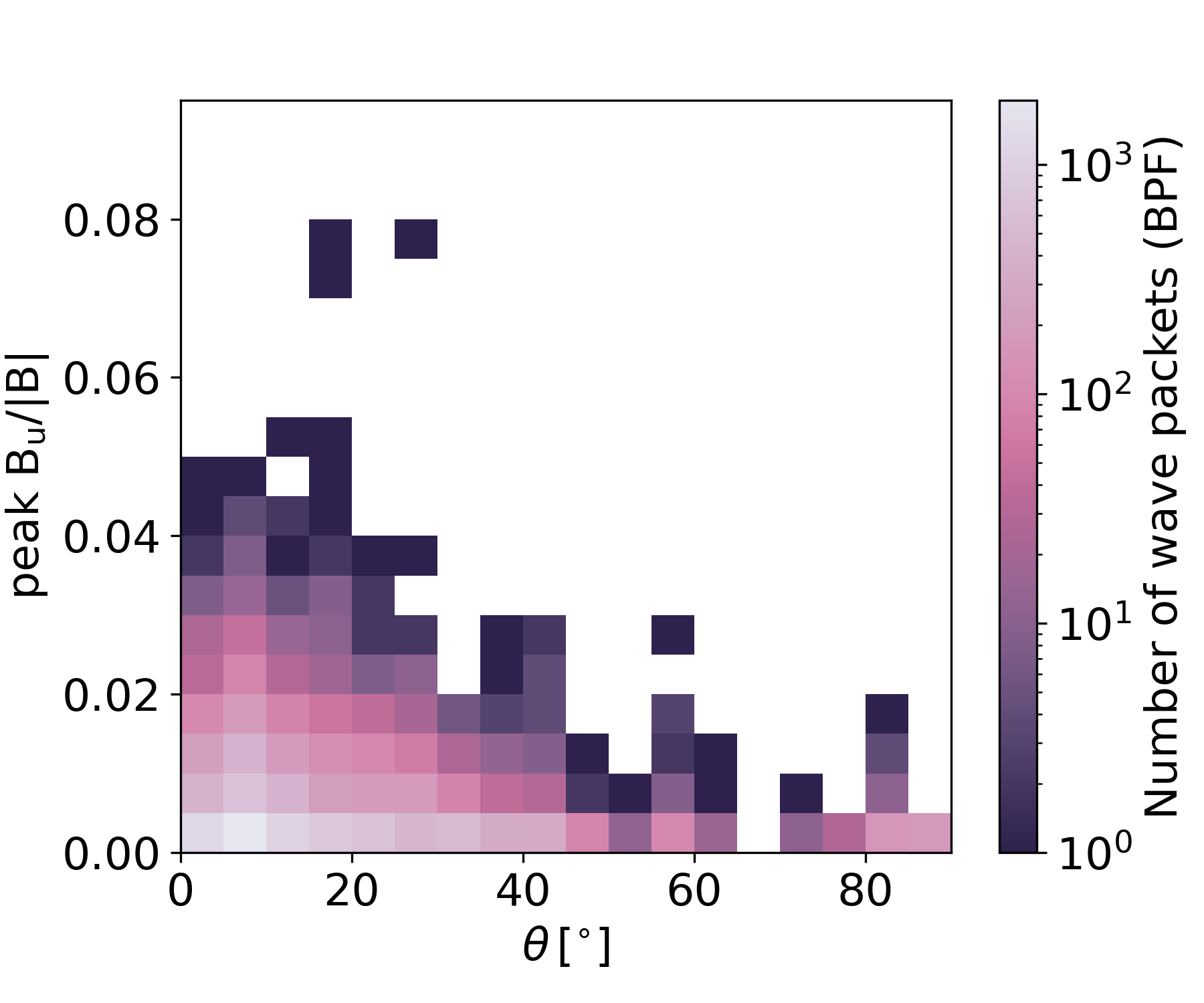} 
     &  	\includegraphics[width=0.45\linewidth]{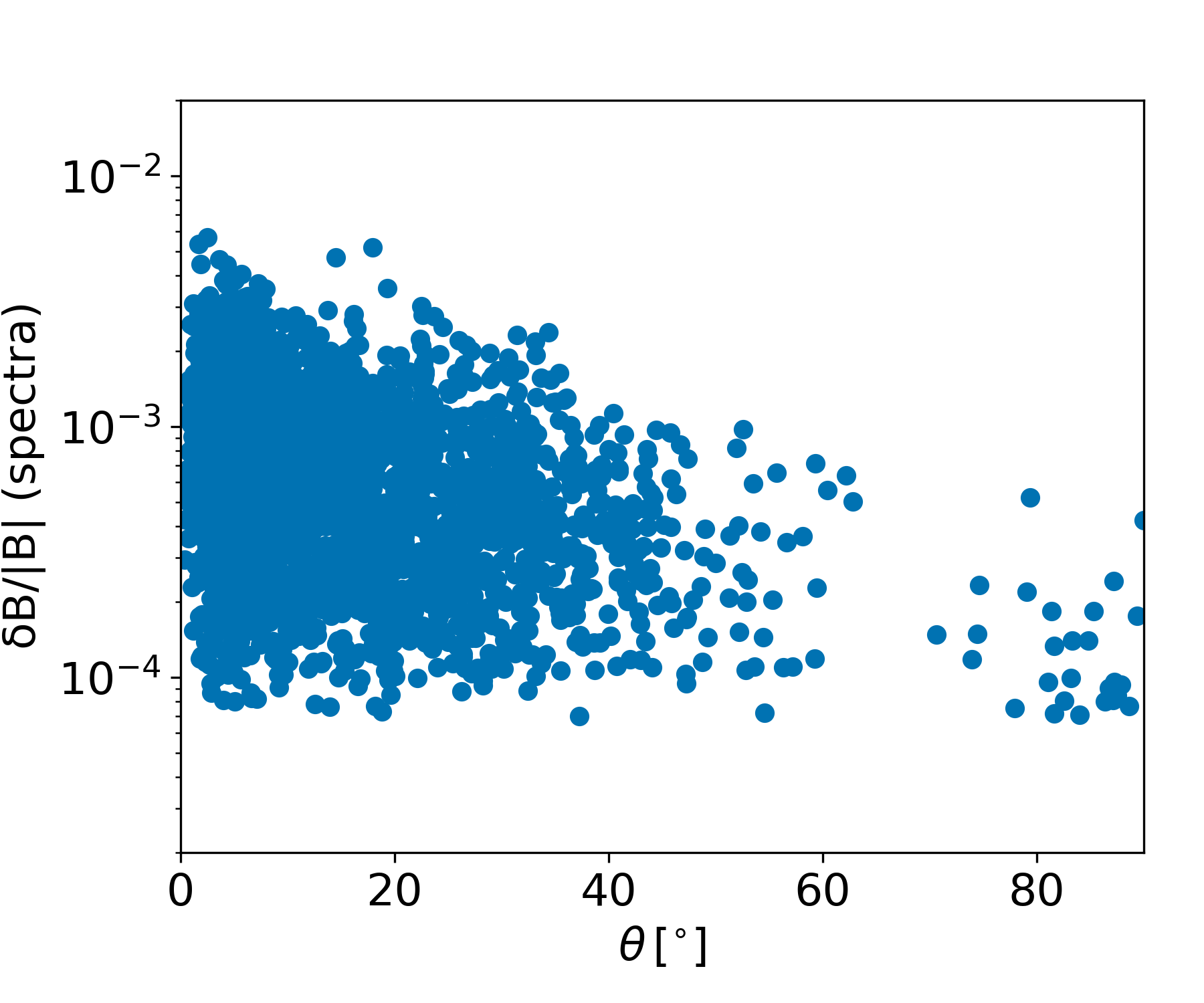}\\
    \end{tabular}
	\caption{Whistler wave amplitudes during the first encounter of PSP (November 2018) from the DFB cross-spectral and BPF measurements compared to the background magnetic field from MAG. \textit{Right:} 2D histogram of the waves relative amplitude and WNA from the BPF and cross-spectral measurements (8481 wave packets, that means BPF bins, in total). \textit{Left:} Relative amplitude of the magnetic fluctuations from the cross spectra versus the WNA for the whistlers (2701 wave packets in total).}
	\label{fig:stat_amp}
\end{figure*}

\subsection{Wave normal angle}\label{sec:wave_normal_angle}

Figure~\ref{fig:stat_theta} gives the histogram of the WNA for all the 2701 wave packets. Most of the whistlers are quasi-parallel. This is consistent with recent studies with Solar Orbiter data \citep{kretzschmar_whistler_2021} and the analysis of the same dataset by \citet{cattell_parker_2022}.
However, a significant part of the whistlers, i.e. 3\%, have WNAs above 45\deg. We notice a depletion of whistlers around the Gendrin angle which is about 66\deg for $0.2f/f_{ce}$ \citep[also observed in magnetospheric studies, e.g.][]{agapitov_synthetic_2018}.
The detailed analyses in Section~\ref{sec:detailed_cases} showed that quasi-parallel whistler wave packets as observed in the cross spectra can show a small proportion of oblique whistlers. Moreover, the oblique cases usually have lower planarity values (45\% of the oblique cases have planarity below 0.65, i.e. near our threshold on the planarity). This was the case for event~3 in particular for which the planarity of the oblique whistlers was found to be below our threshold. We thus conclude that the proportion of oblique whistlers in our statistics is likely a lower limit or at least that a few percent of oblique whistlers may always be associated with quasi-parallel whistlers. We notice that the oblique whistlers have lower frequencies than the quasi-parallel ones, even though the two distributions have similar median values (around \SI{140}{Hz}), the maximum frequency reached by the oblique whistlers is \SI{284}{Hz} compared \SI{531}{Hz} for the quasi-parallel whistlers. In terms of frequencies compared to $f_{ce}$, we found that the oblique whistlers have always frequencies below $0.15 f_{ce}$.

In Figure~\ref{fig:stat_amp} we show the relative amplitude of the whistlers, estimated in two different ways as a function of the WNA. Two different methods are used so the relative amplitude obtained can be compared on the one hand with the amplitude derived from the waveform analysis in Section~\ref{sec:scatter_amp_theta} and on the other hand with other studies where integrated amplitudes are used. For the scatter plot on the right, the relative amplitude is computed as $\delta \mathrm{B}/|\mathrm{B}| = \sqrt{\mathrm{PSD \times \Delta f }}/|\mathrm{B}|$, using the trace of the cross-spectral matrix. The same method was used with the PSD estimated from the waveforms in Section~\ref{sec:scatter_amp_theta}. The wave amplitudes are systematically lower for higher WNA, so that the oblique whistlers amplitudes median values being $2 \times 10^4$ and compared to $6 \times 10^4$ for the quasi-parallel waves. On the left of the figure, a 2D histogram of the relative amplitude of the whistlers as a function of the WNA. We define the relative amplitude of the whistlers as the peak value given in the BPF measurements (i.e., in a specific frequency band) with respect to the local magnitude of the background magnetic field. We consider all the BPF bins within one cross-spectral bin and attribute for each of them the corresponding WNA.  This means that the total number of BPF wave packets is higher than the number of cross spectral wave packets. However, we checked that this does not change the proportion of oblique whistlers, and distributions of the wave parameters. We discard the values of peak $\mathrm{B_u}/|\mathrm{B}|$ that are lower than 0.001. Since some whistlers detected in the cross-spectral bin show a very narrow response in time in the BPF data (see e.g. event~2), this arbitrary threshold is meant to dismiss BPF bins that may not be whistlers. Most of the wave packets have a relative amplitude below 0.02 (97\%) which is consistent with studies near 1 AU \citep[e.g.][]{2019ApJ...878...41T}. Only quasi-parallel whistlers have higher amplitudes in our statistics (up to 0.075). This is consistent with other observational studies and theoretical predictions. The oblique whistlers are more electrostatic and will thus be less intense in the magnetic field than the quasi-parallel whistlers \citep{agapitov_statistics_2013, 2016SSRv..200..261A}.

\begin{figure}
    \centering
    \begin{tabular}{c}
    		\includegraphics[width=\linewidth]{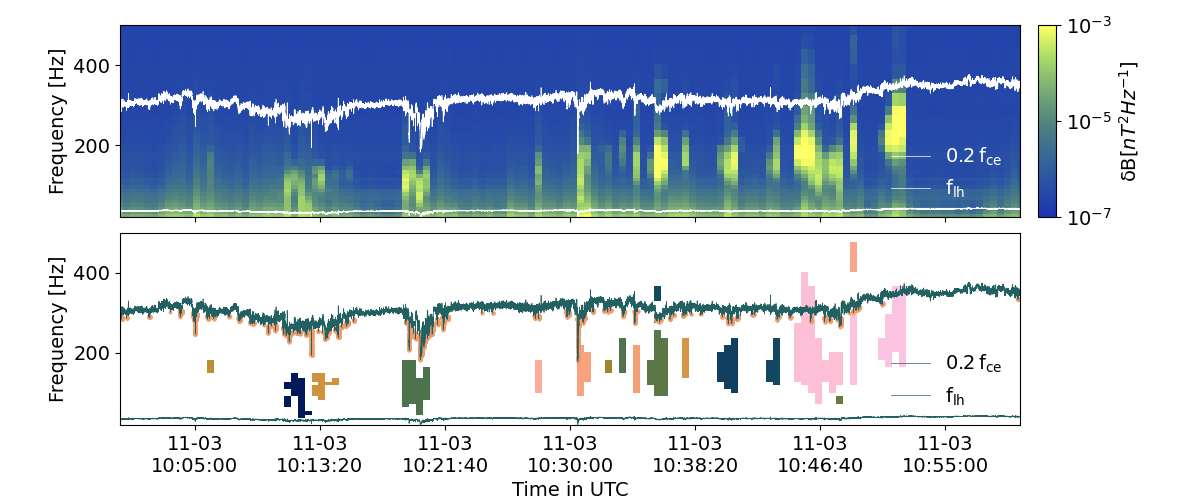} \\
     		\includegraphics[width=\linewidth]{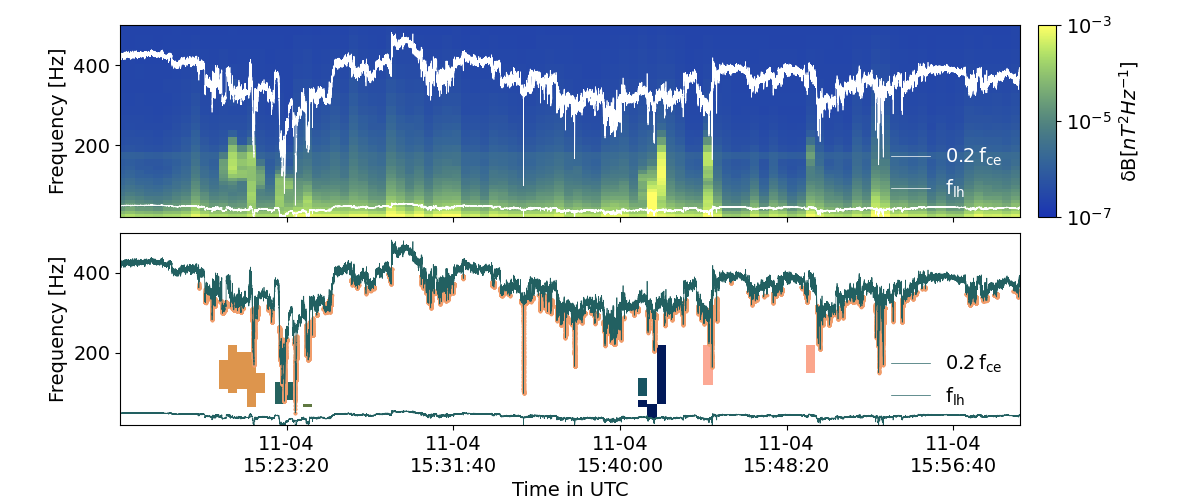}\\
    \end{tabular}
	\caption{Two intervals of whistlers detected in the cross spectra that are collocated with magnetic dips (about one-hour long each). For both figures, the first row shows the trace of the spectral matrices. The two white lines indicate 20\% of the electron-cyclotron frequency $f_{ce}$ and the lower-hybrid frequency $f_{lh}$, respectively. The second row shows the whistler clusters detected, with each a randomly attributed a color.  The two green lines indicate 20\% of the electron-cyclotron frequency $f_{ce}$ and the lower-hybrid frequency $f_{lh}$, respectively. The magnetic dips, detected with the method described in Section~\ref{sec:method_dips}, are highlighted in salmon pink.}
	\label{fig:detections}
\end{figure}

\begin{figure}
    \centering
    \includegraphics[width=\linewidth]{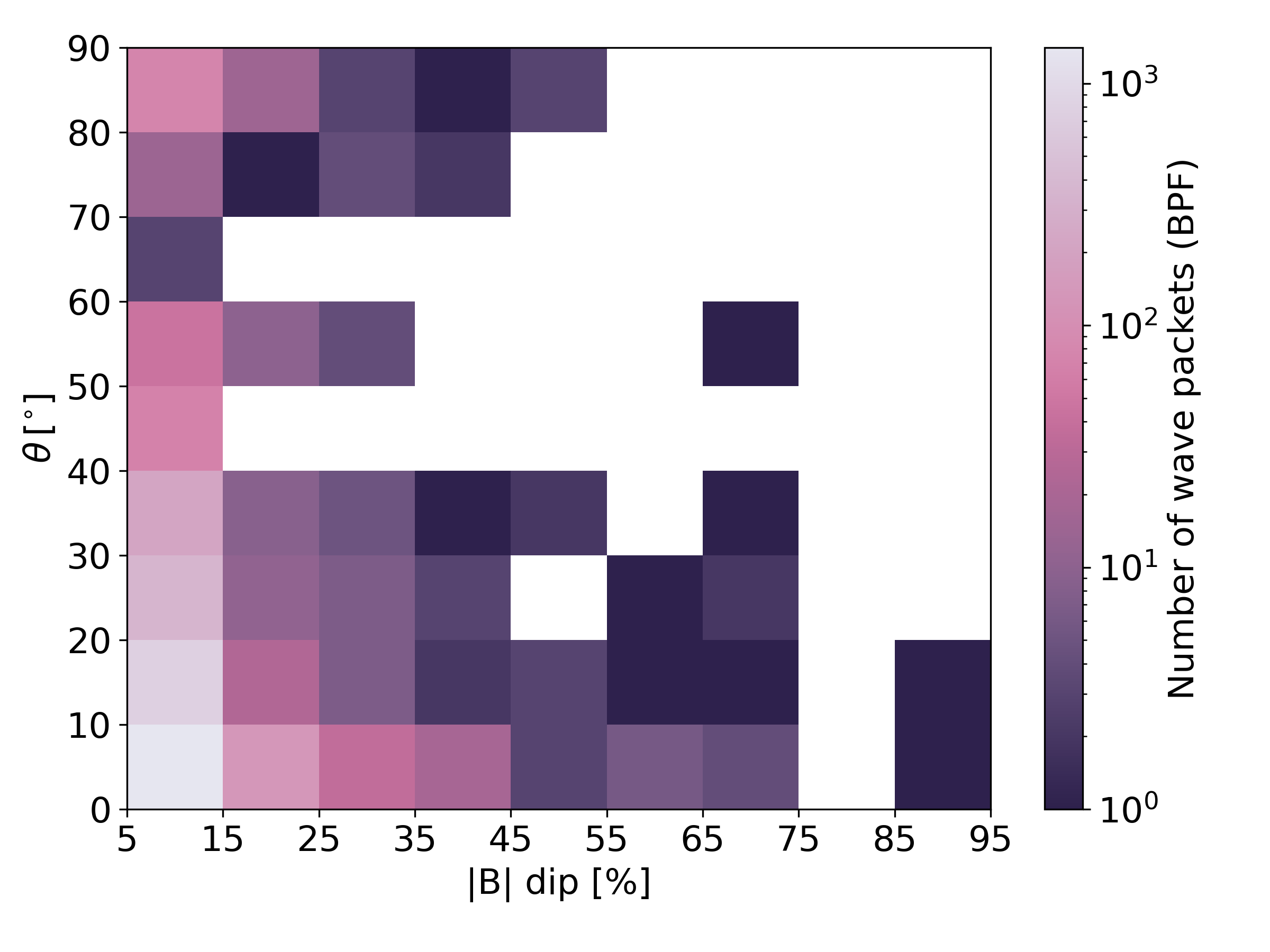} 
	\caption{2D histogram of the WNA of the whistlers and dips in the magnitude of the background magnetic field.}
	\label{fig:stat_dips}
\end{figure}

\subsection{Collocation with magnetic dips}\label{sec:magnetic_dips}

In Figure~\ref{fig:detections}, we highlight two time windows, of about one hour each, for which we detect whistler waves collocated with magnetic dips. For most of the wave clusters, there is at least one magnetic dip. This is not uncommon. We find that 64\% of the wave packets detected in the spectra are collocated with at least one magnetic dip, 69\% when we consider the groups of wave packets that are cotemporal in the spectra, that is with a common timestamp. 
If we look at the quasi-parallel and oblique populations separately, we find that collocation dips are a bit more frequent for the oblique whistlers: 76\% of the oblique wave packets are collocated with dips versus 64\% for the quasi-parallel ones. 
We emphasize that this is a coarse estimation. Our dip detection is not exhaustive. Since we re-interpolate the depth computed (see Section~\ref{sec:method_dips}) to the cadence of the BPF, we do not detect the dips that are shorter than \SI{0.87}{s}.
On top of that, in the dataset we analyzed, wave clusters appear in series, which means a succession of whistler wave clusters, separated by a few minutes, that can last minutes to hours. Some of these storms seem to appear during macro (and shallow) magnetic dips and switchbacks intervals that are not detected by our method.

\subsection{Possible mechanism(s) of formation in magnetic dips}

Most of the dips detected are drops in magnetic field magnitudes less than 15\% of the magnetic field magnitude (5\% being our detection limit). The duration of magnetic field depletion is usually a few seconds. Longer magnetic dips usually contain several elements - a series of overlapping magnetic dips. As we can see in Figure~\ref{fig:stat_dips} that shows a 2D histogram of the WNA versus the drop in the magnetic field magnitude, this kind of small dip is the most common for both quasi-parallel and oblique whistlers. Very large dips in $|\mathrm{B}|$ are quite rare and often contain quasi-parallel whistlers. The local depletion of magnetic field magnitude is the statistical attribute of switchbacks boundaries \citep[]{farrell_magnetic_2020, froment_direct_2021, rasca_magnetic_2022}, which are presumably generated during switchback generation \citep[]{drake_switchbacks_2021}. From event~1 and 2, we see that magnetic dips are present at the boundaries of the switchbacks but can also be present outside these structures.   There is no plasma density increase observed inside the magnetic dips, which suggests the existence of a hot plasma population inside the magnetic field depletion. Such a population supports the pressure balance and can be naturally filtered in during the formation of the structure. Then this population can seed wave generation statistically associated with switchback boundaries at 35-40 solar radii \citep[]{larosa_switchbacks_2021}.
Following the discussion in \citet{agapitov20}, we thus presume that the whistler waves detected inside magnetic dips were generated locally inside the dips by the thermal anisotropy as quasi-parallel and gained obliqueness by propagating to regions with higher magnetic field magnitude and probably different magnetic field direction \citep[similarly to what can be found in the magnetosphere][]{agapitov_statistics_2013}.

Moreover, by analyzing the distribution in frequencies for the whistlers collocated with dips or not, we find that the whistlers collocated with dips tend to have lower frequencies. Indeed, 83\% of the whistlers collocated with dips have frequencies below \SI{200}{Hz}, compared to 56\% for the whistlers without any dips. This could indicate sunward propagation, for at least some of these waves, causing Doppler-shift toward lower frequencies than the frequency in the plasma frame.

\section{Wave-particle interaction perspectives}\label{sec:discussion}

Whistler waves are presumed to be responsible for the enhanced pitch-angle scattering of the super-thermal electron population of the solar wind - the strahl \citep{pagel_scattering_2007, cattell_parker_2021, jagarlamudi_whistler_2021}. However, most of the waves reported at heliocentric distances above 50 solar radii have an anti-sunward propagation \citep[e.g.][]{lacombe_whistler_2014, 2019ApJ...878...41T, kretzschmar_whistler_2021} making them about an order less efficient for strahl scattering in comparison with the sunward propagating waves \citep{verscharen_self-induced_2019}. This is true for the quasi-parallel waves but oblique WNA of whistler waves can increase the scattering efficiency for the anti-sunward propagating waves. The frequent occurrence of oblique WNA in the whistler statistics and in particular at the boundary of switchbacks, which appear to be the regular ingredient of the young solar wind, can significantly contribute to scattering of the strahl population into the halo and modulation of the electron heat flux. Statistical connection of whistler waves with the gradients of the background magnetic field magnitude provides favorable conditions for nonlinear trapping and gyrosurfing acceleration of electrons with energies from 50 eV to 1 keV \citep[]{kis_gyrosurfing_2013, artemyev_cyclotron_2013} that corresponds to the strahl electrons energy range \citep[]{halekas_electrons_2020}. Shorter-lived localized whistler bursts in the magnetic holes could therefore tend to scatter the strahl more efficiently. 
The generation of whistler waves in magnetic field dips is presumably caused by the efficient interaction/damping of the waves on the edges of the magnetic dips through interactions with the strahl - the waves are damped locally around their generation regions. 

\begin{figure*}
    \centering
	\includegraphics[width=\linewidth]{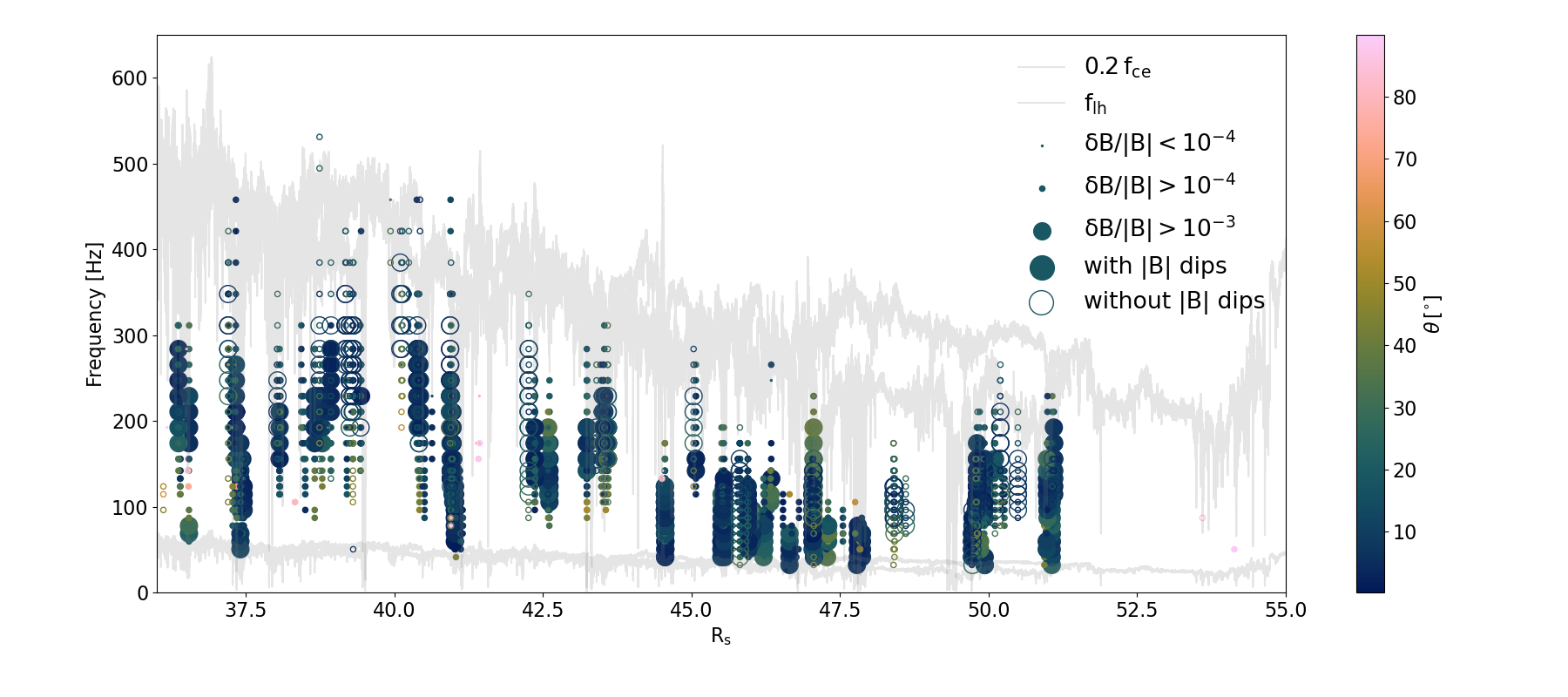} 
	\caption{Summary figure of the properties extracted from the whistlers detected in the DFB cross spectra for PSP encounter~1. Wave frequency versus solar distance. Each dot represents a wave packet in the spectra, with their WNA color-coded, their relative amplitude constraining the size of the dot, and the absence of collocation with $|\mathrm{B}|$ dip resulting in an empty dot. The two gray lines indicate 20\% of the electron-cyclotron frequency $f_{ce}$ and the lower-hybrid frequency $f_{lh}$, respectively, from November 1, 2018, until November 11, 2018. As seen in Section~\ref{sec:global_stat}, most of the whistlers are detected from  November 1, to  November 5. The second part of these gray lines (the outbound phase of the encounter) are thus relevant for very few events.}
	\label{fig:summary_events}
\end{figure*}

\section{Summary}\label{sec:conclusion}

Characterizing whistler wave properties in the solar wind is at the heart of understanding the dynamics and evolution of the eVDF that carries the heat flux. Our study has two main objectives: to shed light on whistler wave properties in the young solar wind by using magnetic field data that cover the appropriate range of frequencies, and validate the statistics we present by studying in detail a few examples that are cross-validated with complementary data sets in terms of cadence, spatial component measured, and frequency and time coverage (cross spectra, BPF, waveforms). 

In Section~\ref{sec:results} we demonstrate that even though the cross spectra have a low cadence of 28 seconds during encounter 1, which is very long compared to typical timescale variations of the background magnetic field, we can derive meaningful statistics of the WNA. Our method relies on the use of BPF measurements that cannot be used directly to derive the WNA but offer a much higher time resolution of 0.87 seconds. This allows us to select the relevant background magnetic field vector associated with the whistler waves. The three events we present in detail show a variety of wave parameters: quasi-parallel to oblique whistlers, and cover different magnetic configurations: boundaries of switchbacks, collocation with magnetic dips or calm intervals. We find that wave normal angles derived from the spectra are in general accordance with the ones derived from the waveforms. We however note that a non-negligible percentage of oblique whistlers can be present in the waveform but hidden in the cross spectra. This is demonstrated by the case of event~1 which shows up fully quasi-parallel from the cross spectra, and event~3 that shows oblique whistlers in the lower frequency band of the event which would be washed out from our statistics due to their low planarity that falls below our threshold.

The properties of the whistlers derived from the cross spectra and presented in Section~\ref{sec:stats} are summarized in Figure~\ref{fig:summary_events}.
The main features are:
\begin{itemize}
    \item Most of the whistler wave packets are quasi-parallel (97\%) to the background magnetic field. We however note that the 3\% fraction of oblique whistlers is likely a lower limit, as revealed by the detailed analysis of Section~\ref{sec:results};
    \item The oblique whistlers tend to have lower frequencies than the quasi-parallel whistlers. Figure~\ref{fig:summary_events} shows that the oblique whistlers are either more narrow band in frequencies than the quasi-parallel whistlers or correspond to the lower frequency band of broader-band wave clusters;
    \item In the observational range of the first encounter of PSP with the Sun, that is between 35 to 55 solar radii, there is no radial dependency of the relative amplitude of the whistlers. This is consistent with the results of \citet{cattell_parker_2022} using BPF measurements on the encounter 1 to 9. However, we note that the radial evolution (of the relative amplitude and other characteristics) should be disentangled from the changes in the solar wind properties in order to fully conclude. We also notice that the oblique whistlers have predictably lower relative amplitudes than the quasi-parallel waves; 
    \item The whistler waves, both quasi-parallel and oblique waves, were often collocated with short-lived magnetic dips (more than 5\% decrease of background magnetic field). This observation supports a possible generation of whistlers in these structures. These waves tend to be detected at lower frequencies than the waves that are not collocated with magnetic dips. This seem could be an indication of sunward propagation and be consistent with a collocation in dips at the boundary of switchbacks \citep{agapitov20}. 
    
\end{itemize}

Whistler waves can efficiently scatter the strahl. Significant broadening of the strahl was observed at the same time as whistlers for this perihelion in \citet{cattell_parker_2021, jagarlamudi_whistler_2021}. In the present paper, we further show that the general properties of most of the detected whistler waves support their generation in magnetic dips. The gradients of the background magnetic field magnitude provides favorable conditions for nonlinear trapping and gyrosurfing acceleration of electrons at energies relevant to the strahl.
These magnetic dips are often found at the boundaries of switchbacks. The occurrence of whistlers in the young solar wind could thus be intimately linked to the occurrence of switchbacks. 
Interestingly, we note that \citet{rasca_magnetic_2022} recently showed that the presence of magnetic dips at switchbacks boundaries is often correlated with the presence of Langmuir waves. \citet{jagarlamudi_whistler_2021}, studying the same encounter as \citet{rasca_magnetic_2022}, showed that Langmuir waves are often present when whistler waves are detected (85~\% of the time). Further work would be needed to understand the relationship between the occurrence of whistlers and Langmuir waves, but it may be that the presence of magnetic dips could favor both types of waves.

Switchbacks are ubiquitous in the young solar wind as measured by PSP \citep{bale19, kasper19, ddw20}. We thus conjecture that magnetic dips are frequent enough to play a significant role in producing the whistlers, beyond the data analyzed in the present paper. This is also supported by the numerical simulations of interchange reconnection of \citet{drake_switchbacks_2021} and \citet{agapitov_flux_2022}. These simulations have shown that magnetic dips can be naturally generated during switchback generation and propagation. However, the presence of magnetic dips are likely not a sufficient condition to the generation of whistlers waves. A low bulk solar wind velocity seem to also be an important condition for the generation of whistlers, as discussed in \citet{jagarlamudi_whistler_2020} and observed in \citet{jagarlamudi_whistler_2020, jagarlamudi_whistler_2021} and in the present paper. Different solar wind conditions may explain why there is a quasi absence of whistlers in the innermost heliosphere (below 28 solar radii) \citep{cattell_parker_2022} where magnetic dips do occur.

Finally, we would like to emphasize that we think results are not in opposition with previous studies of other potential instability mechanisms \citep[i.e. beta-heat flux occurrence consistent with the fan instability in][]{jagarlamudi_whistler_2021, cattell_parker_2022},  different generation mechanisms can cohabit in the young solar wind. We rather highlight that the generation of whistlers waves in magnetic dips in the solar wind may be frequent and should be further investigated in order to understand its impact on the solar wind electron populations.

\begin{appendix}
\section{Cross-spectra frame}\label{sec:appendix}
The $k$-vectors derived from the cross-spectra are rotated first in the SCM frame using the following transformation matrix, and then in the spacecraft frame:
\begin{equation}
    \left( 
    \begin{array}{c}
    \mathsf{k_u} \\
    \mathsf{k_v} \\
    \mathsf{k_w} 
    \end{array}
    \right)
    = \mathsf{R} 
    \left(
    \begin{array}{c}
    \mathsf{k_d} \\
    \mathsf{k_e} \\
    \mathsf{k_f}
    \end{array}
    \right)
\end{equation}
with
\begin{equation}
    \mathsf{R} =
    \left(
    \begin{array}{ccc}
       \ \ 0.4683   &  -0.8134   & 0.3451 \\
         -0.6692      &  -0.0715   & \ \ 0.7396 \\
       -0.5769  &  -0.5773   & -0.5778
    \end{array}
    \right)
    \label{eq_rotation}
\end{equation}

\end{appendix}

\begin{acknowledgements}
C.F., V.K., T.D., L.C., and M.K acknowledge funding from the CNES. V.K. and O.V.A. were supported by NASA grant 80NSSC20K0697 and 80NSSC20K0697; A.L. was supported by STFC Consolidated Grant ST/T00018X/1; O.V.A and S.K. were partially supported by NSF grant number 1914670,  NASA’s Living with a Star (LWS) program (contract 80NSSC20K0218), and NASA grants contracts 80NNSC19K0848, 80NSSC22K0433, 80NSSC22K0522. S. D. B. acknowledges the support of the Leverhulme Trust Visiting Professorship program. V.K.J acknowledges support from the Parker Solar Probe mission as part of NASA's Living with a Star (LWS) program under contract NNN06AA01C. Parker Solar Probe was designed, built, and is now operated by the Johns Hopkins Applied Physics Laboratory as part of NASA’s Living with a Star (LWS) program (contract NNN06AA01C). Support from the LWS management and technical team has played a critical role in the success of the Parker Solar Probe mission. We thank the FIELDS team for providing data (PI: Stuart D. Bale, UC Berkeley). The data used in this work are available on the public data archive NASA CDAWeb (https://cdaweb.gsfc.nasa.gov/index.html/). We also acknowledge the use of the CDPP archive. The authors acknowledge CNES (Centre National d Etudes Spatiales), CNRS (Centre National de la Recherche Scientifique), the Observatoire de PARIS, NASA and the FIELDS/RFS team for their support to the PSP/SQTN data production, and the CDPP (Centre de Donn\'ees de la Physique des Plasmas) for their archiving and provision. The FIELDS experiment on the PSP spacecraft was designed and developed under NASA contract NNN06AA01C. The following acknowledgements were compiled using the Astronomy Acknowledgement Generator (https://astrofrog.github.io/acknowledgment-generator/). Figures were produced using Matplotlib \citep{Hunter:2007} and universally readable color maps \citep{Crameri_2018}.
\end{acknowledgements}

\bibliographystyle{aa}
\bibliography{paper_whistlers}

\end{document}